\documentclass[review]{elsarticle}

\usepackage{lineno,hyperref,amsmath,amssymb,tikz}
\newenvironment{lcases}
  {\left\lbrace\begin{aligned}}
  {\end{aligned}\right.}
\newcommand*\circled[1]{\tikz[baseline=(char.base)]{
            \node[shape=circle,draw,dashed,inner sep=2pt] (char) {#1};}}
\makeatletter
\newcommand{\doublewidetilde}[1]{{%
  \mathpalette\double@widetilde{#1}%
}}
 \newcommand{\invcomma}[1]{``#1''}
\newcommand{\double@widetilde}[2]{%
  \sbox\z@{$\m@th#1\widetilde{#2}$}%
  \ht\z@=.9\ht\z@
  \widetilde{\box\z@}%
}
\makeatother            

\begin{document}

\begin{frontmatter}

\title{On the metric structure of time in classical and quantum mechanics}
%\tnotetext[mytitlenote]{Fully documented templates are available in the elsarticle package on \href{http://www.ctan.org/tex-archive/macros/latex/%contrib/elsarticle}{CTAN}.}

%%%%%%%%%%%%%%%%%%%%%%%%%%%%%%%%%%%%%%%%%%%%%%%%%%%
% Emails
%%%%%%%%%%%%%%%%%%%%%%%%%%%%%%%%%%%%%%%%%%%%%%%%%%%
%\cortext[emailEnrico]{ecattar@ts.infn.it}
%\cortext[emailEnnio]{ennio.gozzi@gmail.com}
%\cortext[emailDanilo]{dmauro75@gmail.com}
%%%%%%%%%%%%%%%%%%%%%%%%%%%%%%%%%%%%%%%%%%%%%%%%%%%
% Authors
%%%%%%%%%%%%%%%%%%%%%%%%%%%%%%%%%%%%%%%%%%%%%%%%%%%
%%%%%%%%%%%%%%%%%%%%%%%%%%%%%%%%%%%%%%%%%%%%%%%%%%%
\author[CommonFirstAddress]{E. Cattaruzza}
\author[CommonFirstAddress,EnnioFirstAddress]{E. Gozzi}
\author[CommonFirstAddress]{D. Mauro}
%%%%%%%%%%%%%%%%%%%%%%%%%%%%%%%%%%%%%%%%%%%%%%%%%%%
% Addresses
%%%%%%%%%%%%%%%%%%%%%%%%%%%%%%%%%%%%%%%%%%%%%%%%%%%
\address[CommonFirstAddress]{INFN, Section of Trieste, \\Via Valerio 2, Trieste, 34100, Italy}
\address[EnnioFirstAddress]{Phys. Dept. Theoretical Section, Univ. of Trieste,\\
 Strada Costiera 11, Miramare, Grignano\\
 Trieste, 34152, Italy}

\begin{abstract}
In this paper we show that, via an extension of time, some metric structures naturally appear in both classical and quantum mechanics when both are formulated via path integrals. We calculate the various Ricci scalar and curvatures associated to these metrics and prove that they can be choosen to be zero in classical mechanics while this is not possible in quantum mechanics. 
\end{abstract}

\begin{keyword}
\PACS 31.15.xk \sep 03.65.-w \sep 04.20.q
\end{keyword}

\end{frontmatter}

%\linenumbers

\section{Introduction}
Quantum gravity is for sure one of the most outstanding open problem in theoretical physics. The usual approach is to take a geometric classical theory 
(like Einstein gravity, string or similar) and apply quantum mechanics to it. People have never reversed the problem, that means first try to understand if there is some hidden universal geometry in quantum mechanics and, second, see if this geometry is compatible with the geometry of the classical model that we want to quantize. In this paper we will concentrate on the first of the two issues above that means try to understand if there is some hidden geometry in quantum mechanics. In doing this we will discover some nice things which may have some application. In a future paper we hope to come back to the second issue mentioned above.  \\
It was shown in \citep{REF1} that classical mechanics (CM) and quantum mechanics (QM) could have a very similar formulation via path-integrals. The generating function of the first, which we will indicate with the acronym $CPI$ (for {\it classical path integral}) has the form:  
\begin{equation}
\mathcal{Z}_{CPI}\left[\mathbb{J} \right]=\int\mathcal D\Phi^a\exp\left[i\int idt d\theta d\bar\theta\,\circled{1}\left(L[\Phi]+\mathbb{J}\,\Phi\right)\right]
\label{eq:1}
\end{equation}
where $\theta,\bar \theta$ are two grassmanian partners of $t$, $\Phi^a$ are  extensions of the phase space coordinates $\varphi^a\equiv(q^1,\dots,q^n;p^1,\dots, p^n),\,a=1,\dots,n$ and $L$ is the usual lagrangian.
The generating functional for quantum mechanics, which we will indicate with $\mathcal Z_{QPI}$ (where QPI stands for {\it quantum path integral}), has the form  
\begin{equation}
\mathcal{Z}_{QPI}\left[\mathbb{J} \right]=\int\mathcal D\Phi^a\exp\left[i\int idt d\theta d\bar\theta\,\circled{$\cfrac{\theta\bar\theta}{\hbar}$}\left(L[\Phi]+\mathbb{J}\,\Phi\right)\right]
\label{eq:2}
\end{equation}
which is very similar to Eq.~(\ref{eq:1}) except that the \circled{1} in Eq.~(\ref{eq:1})  is replaced by \circled{$\frac{\theta\bar\theta}{\hbar}$} in Eqs.~(\ref{eq:2}). 
As these quantities multiply the measure of integration $\int idt d\theta d\bar\theta$, it comes natural to do the following: let us introduce a general dreinbein $E_A^{\,M}$ in the space $t,\,\theta,\bar\theta$ and let us build the following path-integral   
\begin{equation}
\mathcal{Z}_{GPI}\left[\mathbb{J} \right]=\int\mathcal D\Phi^a\exp\left[i\int idt d\theta d\bar\theta\,\circled{$E$}\left(L[\Phi]+\mathbb{J}\,\Phi\right)\right]
\label{eq:3}
\end{equation}
(GPI stands for {\it General Path Integral}) and where $E$ is the determinant (or superdeterminant) of $E_A^M$. Immediately we notice, comparing  Eq.~(\ref{eq:3}) with Eq.~(\ref{eq:1})  and Eq.~(\ref{eq:2}) that the CPI can be considered a GPI with $E=1$ and the QPI a GPI with $E=\frac{\theta\bar\theta}{\hbar}$. As the GPI has a general covariance in the $t,\theta,\bar\theta$ space we could consider the CPI and QPI as two ``gauge fixed'' version of Eq.~(\ref{eq:3}). The reader could object to this by saying that Eq.~(\ref{eq:1}) and  Eq.~(\ref{eq:2}) should contain the Fadeev-Popov determinant if  considered as gauge fixed versions of Eq.~(\ref{eq:3}). The reason we did not put them in   Eq.~(\ref{eq:1}) and Eq.~(\ref{eq:2}) is because we did not path integrate over the variables $E^{A}_{\,B}$ and the gauge-fixing does not depend on the matter field $\Phi^a$. Somehow we can consider the formulation in Eq.~(\ref{eq:1}), Eq.~(\ref{eq:2}), Eq.~(\ref{eq:3}) similar to that of a field theory $\Phi$ in a background gravitational field $E^{A}_{\,B}$. The reader may also object that if both Eq.~(\ref{eq:1}) and Eq.~(\ref{eq:2}) are different gauge of Eq.~(\ref{eq:3}) then we could turn classical mecnhainics Eq.~(\ref{eq:1}) into quantum mechanics Eq.~(\ref{eq:2}) via a general covariant transformation in the extension of time. We will show later on why it is not possible to turn CM into QM Next we will prove that there are various families of $E^{A}_{\,B}$ which give the same CPI and the same for the QPI. These families are parametrized by 4 parameters for the CPI and by 5 for the QPI. From the $E^{A}_{\,B}$ with the help of {\it Wolfram Mathematica} we will build the metric, the Christoffel symbols, the Ricci curvature tensor $R_{\alpha\beta}$ and the Ricci scalar $R$. All of these depend on the same parameters as the $E^{A}_{\,B}$. For the CPI we prove that there is a point in parameter space for which the Ricci scalar and tensors are zero. The same does not happen for the QPI. This fact may indicate something very profound but we have not been able to investigate it.  We leave to the reader the task of further explore this last issue.
The paper is organized as follows. In Section~(\ref{Section:2}) for completness we briefly review (\ref{eq:1}) and  (\ref{eq:2}). In Section~(\ref{Section:3})  we introduce the vierbein $E^{A}_{\,B}$ and indicate the general strategy. In Section~(\ref{Section:4}) we show how to obtain the vierbein for both the CPI and the QPI and do the counting of the free parameters. In Section~(\ref{Section:5}) we calculate 
we build the metric for both the CPI and the QPI. In Section~(\ref{Section:6}) we proceed to calculate the Ricci scalar curvature for both theories. In Section~(\ref{Section:7}) we search for the point in parameter space where the Ricci scalar and tensors are zero in the CPI. We also prove that a similar point does not exist for the QPI. In Section~(\ref{Section:8}) we summarize what we had done and the prospects for the future. In few appendices we confine some detailed and long calculations.
\section{Review\label{Section:2}}
In the thirties Koopman and Von Neumann (KVN) proposed \citep{REF2.1,REF2.2} an operatorial and Hilbert space formulation of classical mechanics (CM) on the lines of what had been done few years before for quantum mechanics (QM). It was then natural to give \citep{REF3.1,REF3.2,REF3.3,REF3.4,REF3.5,REF3.6,REF3.7,REF3.8,REF3.9,REF3.10,REF3.11} a path-integral version of the KVN formalism like Feynman had done for the operatorial version of QM \citep{REF4}. Actually, the path-integral version of classical mechanics (CPI) provided in a natural way a generalization of the KVN formalism in  the sense that it gave the classical evolution of differential forms and tensors on phase-space \citep{REF5}.   \\ The procedure has been worked out in details in \citep{REF1} and \citep{REF3.1,REF3.2,REF3.3,REF3.4,REF3.5,REF3.6,REF3.7,REF3.8,REF3.9,REF3.10,REF3.11} and can be summarized as follows. The KVN postulates for the Hilbert space and operatorial versioon of CM are the following:
\begin{enumerate}
\item a state of a classical system, whose phase-space is indicate by $\mathcal M$ with coordinates $\varphi^a\equiv(q^1,\dots,q^n;p^1,\dots, p^n)$  is represented by an elememt $|\psi\rangle $ of the Hilbert space $\mathcal H$.
\item On this Hilbert space the operators $\hat p^i$ and $\hat q^j$, whose eigenvalues are $p^i$ and $q^j$, commutes 
\[
[\hat p^i,\hat q^j]=0
\] and their common eigenstates are indicated as $|q,p\rangle$.
\item The staes $\langle q,p|\psi\rangle$ are square-integrable and their modulus squared $|\psi(q,p)|^2$ is the probability density $\rho(q,p)$ of finding the system in $(q,p)$.
\item The evolution of $\psi(q,p)$ is given by the Liouville equation
\[
i\frac{\partial \psi}{\partial t}=\widehat L \psi
\]
where the Liouvillian $\widehat{L}$ is 
\[
\widehat{L}=i\,\left(\frac{\partial H}{\partial q}\,\frac{\partial \,}{\partial p}-\frac{\partial H}{\partial p}\,\frac{\partial \,}{\partial q}\right)
\]
\end{enumerate}
and $H$ is the Hamiltonian of the sistem whose associated equation of motion are 
\begin{equation}
\dot\varphi^a=\omega^{ab}\,\frac{\partial H}{\partial \varphi^b} 
\label{eq:11-1}
\end{equation}
with 
\[
\omega^{ab}=\left(
\begin{matrix}
0 & \mathbb I\\
-\mathbb I & 0
\end{matrix}
\right)
\]
a $2n\times 2n$ matrix called symplectic matrix. It is well-known that the evolution between some initial 
point $\varphi_i$ and some final point  $\varphi_j$ has the following form on the states
\begin{equation}
\psi(\varphi_f,t_f)=\int \mathcal K(\varphi_f,t_f|\varphi_i,t_i)\,\psi(\varphi_i,t_i)\,d^{2n}\varphi_i
\label{eq:12-1}
\end{equation}
where 
\[
\mathcal K(\varphi_f,t_f|\varphi_i,t_i)=\delta[\varphi_f-\Phi_{cl}(t_f;q_i,t_i)]
\]
with $\Phi_{cl}$  the solution of Eq.~(\ref{eq:11-1}) with initial condition $\varphi_i$.
Slicing the time interval $t_f-t_i$ in $N$ intervals, we can re-write the kernel $K(\varphi_f,t_f|\varphi_i,t_i)$ as follows:
\begin{equation}
K(\varphi_f,t_f|\varphi_i,t_i)=\displaystyle\lim_{N\to\infty}\left\{\int \prod^{N-1}_{J=1} d\varphi_J\,\delta[\varphi_J-\Phi_{cl}(t_J;\varphi_i,t_i]\right\}\,\delta[\varphi_f-\Phi_{cl}(t_f;q_i,t_i)]
\label{eq:12-2}
\end{equation}
where $\varphi_J$ are the intermediate points between $\varphi_i$ and $\varphi_f$ over which we integrate. The Dirac deltas which appear in Eq.~(\ref{eq:12-1}) can be written as 
\begin{equation}
\delta[\varphi^J-\Phi_{cl}(t_I;\varphi_i,t_i)]=\left.\delta\left[\dot\varphi^a-\omega^{ab}\,\frac{\partial H}{\partial \varphi^b}\right]\right|_{t_I}\left.\textrm{det}\left[\delta^a_b\partial_t-\omega^{ac}\frac{\partial^2H}{\partial\varphi^c\partial\varphi^b}\right]\right|_{t_J}.
\label{eq:13-1}
\end{equation}
Let us now introduce some auxiliary variables $\lambda_a$ and let us rewrite the first term on the RHS of Eq.~(\ref{eq:13-1}) as 
\begin{equation}
\delta\left[\dot\varphi^a-\omega^{ab}\,\frac{\partial H}{\partial \varphi^b}\right]=\int d\lambda_a\,\exp\left[i\,\lambda_a\left(\dot\varphi^a-\omega^{ab}\,\frac{\partial H}{\partial \varphi^b}\right)\right]
\label{eq:13-2}
\end{equation}
modulo a normalization factor. 
Let us also introduce $4n$ grassmanian variables \citep{REF6} $c^a,\bar c_a,a=1,\dots,2n$ so that we can rewrite the det on the RHS of Eq.~(\ref{eq:13-1}) as:
\begin{equation}
\textrm{det}\left[\delta^a_b\partial_t- \omega^{ac}\frac{\partial^2H}{\partial\varphi^c\partial\varphi^b}\right]=
\int dc^a d\bar c_a \,\exp\left[-\bar c_a\left(\delta^a_b\partial_t-\omega^{ac}\frac{\partial^2H}{\partial\varphi^c\partial\varphi^b}\right)c^b\right].
\label{eq:14-1}
\end{equation}
Using Eq.~(\ref{eq:13-1}), Eq.~(\ref{eq:13-2}) and Eq.~(\ref{eq:14-1}) in Eq.~(\ref{eq:12-2}) we get
\begin{equation}
K(\varphi_f,t_f|\varphi_i,t_i)=\int_{\varphi_i}^{\varphi_f} \mathcal D''\varphi\mathcal{D}\lambda\mathcal{D}c\mathcal{D}\bar c \,\exp\left[i\int dt \widetilde{\mathcal{L}}\right]
\label{eq:14-2}
\end{equation}
where 
\begin{equation}
\widetilde{\mathcal{L}}=\lambda_a\left[\dot\varphi^a-\omega^{ab}\,\frac{\partial H}{\partial \varphi^b}\right]+i\,\bar c_a\left(\delta^a_b\partial_t-\omega^{ac}\frac{\partial^2H}{\partial\varphi^c\partial\varphi^b}\right)c^b
\label{eq:14-3}
\end{equation}
and $\mathcal{D}''\varphi$ indicates that the integration is done over all its intermediate points and not on the end points $\varphi_i$ and $\varphi_f$.  Eq.~(\ref{eq:14-2}) is basically the path-integral counter-part of the KVN formalism. Remembering   how commutators are obtained from the path-integral \citep{REF4} we get 
\begin{eqnarray}
\label{eq:15-1}
[\hat{\varphi}^a,\hat{\varphi}^b]&=&0\nonumber\\
\lbrack\hat{\varphi}^a,\hat{\lambda}_b\rbrack&=&i\,\delta^a_b\\
\lbrack\hat{\bar c}_a,\hat{c}^b\rbrack&=&\delta^a_b\nonumber
\end{eqnarray}
where the last are anticommutators or graded commutators \citep{REF6}. All the other commutators are zero. From the second commutators of Eq.~(\ref{eq:15-1}) we can realize operatorially the $\hat \lambda_a$ as 
\begin{equation}
\hat \lambda_a=-i\frac{\partial\,}{\partial \varphi^a}. 
\label{eq:15-2}
\end{equation} 
Let us now see how the Liouville operator emerges from Eq.~(\ref{eq:14-2}) from the non-grassmanian part of $\widetilde{\mathcal{L}}$ which we indicate with $\tilde{\mathcal{L}}_B$ the following quantitity:
\[
\widetilde{\mathcal{L}}_B=\lambda_a\,\dot{\varphi}^a-\widetilde{\mathcal{H}}_B
\]
where 
\begin{equation}
\widetilde{\mathcal{H}}_B=\lambda_a\,\omega^{ab}\frac{\partial H}{\partial \varphi^b}.
\label{eq:16-1}
\end{equation}
It is then clear that 
\begin{equation}
\int \mathcal{D}\varphi\mathcal{D}\lambda\,\exp\left[i\int dt \widetilde{\mathcal L}_B\right]\longrightarrow \exp\left[-\widehat{\widetilde{\mathcal{H}}}_B\,t\right]
\label{eq:16-2}
\end{equation}
where $\widehat{\widetilde{\mathcal{H}}}_B$ is the operator associated to Eq.~(\ref{eq:16-1}) obtained using Eq.~(\ref{eq:15-2})
\begin{eqnarray}
{\widetilde{\mathcal{H}}}_B&=&-i\frac{\partial}{\partial \varphi^a}\omega^{ab}\frac{\partial H}{\partial \varphi^b}=\nonumber \\
&=&-i\frac{\partial H}{\partial p} \frac{\partial}{\partial q}+i\frac{\partial H}{\partial q} \frac{\partial}{\partial p}=\widehat{L}
\end{eqnarray}
and this  $\widehat{L}$ is the Liouville operator. 
The reader may ask now which operator we would get if we had kept also the grassmanian variables. It was shown in \citep{REF3.1} that the $c^a$ can be identified with the differential operator $d\varphi^a$. Via these we can build generic differential forms \citep{REF5} 
\begin{equation}
\psi(\varphi,d\varphi)
\label{eq:17-1}
\end{equation}
and we know that their evolution is given by an operator \citep{REF5} called the {\it Lie derivative of the Hamiltonian flow} which is simbolically written as $\mathcal L_{(dH)^{\#}}$. So 
\begin{equation}
\partial _t\psi(\varphi,d\varphi)=\mathcal L_{(dH)^{\#}}\psi(\varphi,d\varphi).
\label{eq:17-2}
\end{equation}
This operator is a generalization of the Liouville operator which makes the evolution of $\psi(\varphi)$ which are called zero forms. As we said  in  \citep{REF3.1,REF3.2,REF3.3,REF3.4,REF3.5,REF3.6,REF3.7,REF3.8,REF3.9,REF3.10,REF3.11} the $c^a$ of our path-integral can be identified with the $d\varphi^a$ and so the differential form Eq.~(\ref{eq:17-1}) can be turned into a $\psi(\varphi,c)$
\begin{equation}
\psi(\varphi,d\varphi)\longrightarrow \psi(\varphi,c).
\label{eq:18-1}
\end{equation}
From the path-integral Eq.~(\ref{eq:14-2}) it is clear that the evolution of $\psi(\varphi,c)$ is given by
\begin{equation}
\partial_t\phi(\varphi,c)=\widehat{\widetilde{H}}\psi(\varphi,c),
\label{eq:18-2}
\end{equation} 
where $\widehat{\widetilde{H}}$ is the operatorial Hamiltonian associated to the Lagrangian Eq.~(\ref{eq:14-3}).
Comparing Eq.~(\ref{eq:18-2}) with Eq.~(\ref{eq:17-2}) we can say that the Hamiltonian operator of our path-integral is a well-known object \citep{REF5} in differential geometry, i.e. it is the {\it Lie derivative of the Hamiltonian flow}. The identification with objects of differential geometry can be also extended to  the exterior derivatives, the inner contractions, the Lie brackets  and the whole Cartan calculus \citep{REF5}.   
The details of this important correspondence have been worked out in \citep{REF3.1,REF3.2}. So the auxiliary variables that we introduced $c^a,\bar c_a,\lambda_a$ are not just tricks to rewrite the path integral in a simpler form but crucial geometrical objects.  Let us now go back to the commutation relations Eq.~(\ref{eq:15-1}). We said before that we can realize the $\hat\lambda_a$ as a derivative operator (like in Eq.~(\ref{eq:15-2}) and obviously the $\varphi ^a$ as a multiplicative one:
\begin{equation}
\hat\varphi ^a|\varphi\rangle= \varphi^a\,|\varphi\rangle.
\label{eq:20-1}
\end{equation}  
The same can be done for the operators $\hat c^a,\hat{\bar c}_a$. As they commute with the $\hat{\varphi}^a$ and $\hat{\lambda}_a$,  we can generalize the states of Eq.~(\ref{eq:20-1}) to the following ones: 
\begin{equation}
\label{eq:20-2}
\left\{
\begin{matrix}
\hat\varphi ^a|\varphi,c\rangle&=& \varphi^a\,|\varphi,c\rangle\\
\hat c ^a|\varphi,c\rangle&=& c^a\,|\varphi,c\rangle\\
\end{matrix}
\right.
\end{equation}
and implement $\hat{\bar{c}}_a$ as a derivative operator 
\[
\hat{\bar{c}}_a=\frac{\partial\,}{\partial c^a}.
\]
There is another manner to realize the Eq.~(\ref{eq:15-1}). Noting that $\hat q^i$ and $\hat \lambda_{p_i}$ commutes and the same $\hat c^q$ and $\hat c_p$ we can diagonalize these operators \footnote{The index $q$ and $p$ on $\hat c,\hat{\bar{c}},\hat{\lambda}$ indicates respectively the first and the last n-indices on $\hat c^a,\hat{\bar{c}}^a,\hat{\lambda}^a$} and obtain the states 
\begin{equation}
\left\{
\begin{matrix}
\hat  q\,|q,\lambda_p,c^q,\bar c_p\rangle&=& q\,|q,\lambda_p,c^q,\bar c_p\rangle\\
\hat  \lambda_p\,|q,\lambda_p,c^q,\bar c_p\rangle&=& \lambda_p\,|q,\lambda_p,c^q,\bar c_p\rangle\\
\hat  c^q\,|q,\lambda_p,c^q,\bar c_p\rangle&=& c^q\,|q,\lambda_p,c^q,\bar c_p\rangle\\
\hat{\bar{c}}_p\,|q,\lambda_p,c^q,\bar c_p\rangle&=& \bar c_p\,|q,\lambda_p,c^q,\bar c_p\rangle\\
\end{matrix}
\right.
\label{eq:21-1}
\end{equation}
while the operators $\hat p$ and $\hat \lambda_q$ are realized as derivatives operators 
\[
\hat p=i\,\frac{\partial\,}{\partial\lambda_p},\quad \hat \lambda_ q=-i\,\frac{\partial\,}{\partial q}.
\]
The two basis Eq.~(\ref{eq:20-2}) and Eq.~(\ref{eq:21-1}) are related by Fourier transformation \citep{REF1}.
The transition amplitudes in the basis of Eq.~(\ref{eq:21-1}) is a generalization of  Eq.~(\ref{eq:14-2})  
and it has the following path-integral expression
\begin{equation}
\langle \varphi_f,c_f,t_f|\varphi_i,c_i,t_i\rangle=\int \mathcal{D}''\varphi\mathcal{D}\lambda\mathcal{D}''c\mathcal{D}\bar c\,\exp\left[i\int dt\widetilde{\mathcal L}\right]
\label{eq:22-1}
\end{equation}
where the integration over $c$ has been limited to the internal points with the end-points fixed and indicated with $\mathcal{D}''c$.
Using the basis of Eq.~(\ref{eq:21-1}) the transition amplitudes will have the path-integral form
\begin{eqnarray}
\label{eq:22-2}
&&\langle q_f,\lambda^p_f,c_f^q,\bar c_f^p|q_f,\lambda^p_i,c_i^q,\bar c_i^p\rangle =\\
&=&\int \mathcal{D}''q\mathcal{D}p\mathcal{D}''\lambda^p\mathcal{D}\lambda^q\mathcal{D}''c^q\mathcal{D}c^p\mathcal{D}\bar c^q\mathcal{D}''\bar c^p\,\exp\left[i\int dt\doublewidetilde{\mathcal L\,}\right]\nonumber
\end{eqnarray}
where $\widetilde{\widetilde{\mathcal L}}$ is a Lagrangian which differ from  $\widetilde{\mathcal{L}}$ of Eq.~(\ref{eq:14-3}) by surface terms. More details can be found in ref.\citep{REF1}. At this point we have to introduce two crucial ingredients which are familiar from the supersymmetry formalism \citep{REF7}. Let us extend the variable $t$ via two grassmanian partners $\theta,\bar\theta$. The triplet $(t,\theta,\bar\theta)$ is often called ``supertime". If we extend $t$ to the 4-dim  $x^{\mu}$ then there is an analog extension of super-time called ``superspace'' \citep{REF7}. With this tool we can group-together the various variables $(\varphi^a,\lambda_a,c^a,\bar c_a)$ into a function of $(t,\theta,\bar\theta)$ called superfield and defined as follows: 
\begin{equation}
\Phi^a(t,\theta,\bar\theta)\equiv \varphi^a+\theta\,c^a+\bar\theta\,\omega^{ab}\,\bar c_b+i\,\bar \theta\theta\,\omega^{ab}\lambda_b.
\label{eq:23-1}
\end{equation}
We can separate off the $q$ and $p$ part of this superfields as follows
\begin{equation}
\Phi^a=\left(
\begin{matrix}
Q_i\\P_i
\end{matrix}
\right)\equiv \left(
\begin{matrix}
q_i\\p_i
\end{matrix}
\right)+\theta \left(
\begin{matrix}
c^{q_i}\\c^{p_i}
\end{matrix}
\right)+\bar\theta\left(
\begin{matrix}
\bar c_{q_i}\\-\bar c_{p_i}
\end{matrix}
\right)+i\bar\theta\theta \left(
\begin{matrix}
\lambda_{p_i}\\-\lambda_{q_i}
\end{matrix}
\right).
\label{eq:24-1}
\end{equation}
Using the superfield there are some nice identities which we will need later on.
Let us build the Lagrangian associated to $H(\varphi)$ of the original equations of motion Eq.~(\ref{eq:11-1})
and let us call it $L(\varphi)$ where we replaced $\dot q$ with $p$. Let us now replace in $L(\varphi)$ the $\varphi$ with the superfield $\Phi^a$ and expand in $\theta,\bar\theta$. We get  
\begin{equation}
L[\Phi]=L(\varphi)+\theta\,\mathcal{M}+\bar{\mathcal{M}}\bar\theta-i\bar\theta\theta\,\doublewidetilde{\mathcal{L}\,}
\label{eq:24-2}
\end{equation}
where $\doublewidetilde{\mathcal{L\,}}$ is the Lagrangian which enters in Eq.~(\ref{eq:22-1}). We will need these identities later on. Let us drop the indices in Eq.~(\ref{eq:24-1}):
\begin{equation}
\left\{
\begin{matrix}
Q(\theta,\bar\theta)&=&q+\theta\,c^q+\bar\theta\,c_p+i\,\bar\theta\theta\,\lambda_p\\
P(\theta,\bar\theta)&=&p+\theta\,c^p-\bar\theta\,c_q-i\,\bar\theta\theta\,\lambda_q
\end{matrix}
\right..
\label{eq:25-1}
\end{equation}
As the variables which enter $Q$ they all commute once they are turned into operators, we could define the following states
\begin{equation}
\widehat Q|Q\rangle=Q(t,\theta,\bar \theta)|Q\rangle
\label{eq:25-2}
\end{equation}
which clearly satisfy 
\[\left\{
\begin{matrix}
\hat q|Q\rangle&=&q\,|Q\rangle\\
\hat \lambda_p|Q\rangle&=&\lambda_p\,|Q\rangle\\
\hat c^q|Q\rangle&=&c^q\,|Q\rangle\\
\hat{\bar c}_p|Q\rangle&=&\bar{c}_p\,|Q\rangle.\\
\end{matrix}
\right.
\label{eq:25-3}
\]
So we can identify the states $|Q\rangle$ with those of the basis Eq.~(\ref{eq:25-2}). We can now use this fact and Eq.~(\ref{eq:24-2})
to rewrite Eq.~(\ref{eq:22-2}) as follows
\begin{equation}
\langle Q_f,t_f|Q_i,t_i\rangle=\int \mathcal{D}''Q\mathcal{D}P\,\exp\left[i\int_{t_0}^t\,i\,dt' d\theta d\bar\theta\,L[\Phi]\right]
\label{eq:26-1}
\end{equation}
 where we have used the standard rule of grassmanian integration 
 \[
\int d\theta d\bar\theta \,\bar\theta\theta = 1. 
 \]
 all the details above are carefully explained in ref-\citep{REF1}.
Let us go back to the quantum mechanical path-integral \citep{REF4} which gives the following expression for the transition amplitude
\begin{equation}
\langle q_f,t_f|q_i,t_i\rangle=\int \mathcal{D}''q\mathcal{D}p\,\exp\left[\frac{i}{\hbar}\int dt' \,L[\varphi]\right].
\label{eq:27-1}
\end{equation}
Note the great analogy between the classical path-integrac (CPI) Eq.~(\ref{eq:26-1}) and the quantum path-integral (QPI) Eq.~(\ref{eq:27-1}). We pass from one to the other by the following steps:
\begin{equation}
\left\{
\begin{matrix}
Q&\longrightarrow & q\\
P&\longrightarrow & p\\
i\int dt d\theta d\bar\theta&\longrightarrow& 1/\hbar
\end{matrix}
\right..
\label{eq:27-2}
\end{equation}
This is a sort of dimensional reduction which in \citep{REF1} we proved to be equivalent to geometric quantization \citep{REF8}. More details can be found in ref.\citep{REF1}. Differently than in \citep{REF1} in this paper we still exploit  the relation between Eq.~(\ref{eq:27-1}) and Eq.~(\ref{eq:26-1}) but following a different route. Let us write Eq.~(\ref{eq:27-1}) using the superfield and relation Eq.~(\ref{eq:24-2}) 
\begin{equation*}
\int dt\,L(\varphi)=\int dt d\theta d\bar\theta \,\bar\theta\theta\,L[\Phi]\\
=\int i\, dt d\theta d\bar\theta \,(-i\,\bar\theta\theta)\,L[\Phi]
\end{equation*}
so 
\[
\frac{i}{\hbar}\int dt\,L(\varphi)=i\,\int i\, dt d\theta d\bar\theta \,\left(-\frac{i}{\hbar}\,\bar\theta\theta\right)\,L[\Phi].
\]
We can then rewrite Eq.~(\ref{eq:27-1}) as 
\begin{equation}
\langle  q_f,t_f|q_i,t_i\rangle
=\mathcal N \int \mathcal{D}''Q\mathcal{D}P\,\exp\left[i\int \,i\,dt' d\theta d\bar\theta \,\left(-\frac{i}{\hbar}\,\bar\theta\theta\right)\,L[\Phi]\right].
\label{eq:28-2}
\end{equation}
where $\mathcal N$ is a normalizing factor. On the right hand side of Eq.~(\ref{eq:28-2}) the integration over $c,\bar c,\lambda$ drops off the path-integration because these variables do not enter the weight. The normalizing factor $\mathcal N$ is there to get $1$ out of those extra intergrations.
We could avoid introducing this normalizing factor  if we write the L.H.S. of  Eq.~(\ref{eq:28-2}) as $\langle Q_f,t_f|Q_i,t_i\rangle$
The pieces $\langle \lambda_f,t_f|\lambda_i,t_i\rangle$ $\langle c_f,t_f|c_i,t_i\rangle$ $\langle \bar c_f,t_f|\bar c_i,t_i\rangle$ turn out to be products of ``1'' at each slice in time exactly as on the path-integral on the R.H.S. So we can summarize the  Eq.~(\ref{eq:28-2}) and  Eq.~(\ref{eq:26-1}) as 
 \begin{equation}
\langle Q_f,t_f|Q_i,t_i\rangle_{CPI}=
\int \mathcal{D}''Q\mathcal{D}P\,\exp\left[i\int_{t_0}^t\,i\,dt' d\theta d\bar\theta\,\circled{$\mathbb I$}\,L[\Phi]\right]
\label{eq:29-1}
 \end{equation}
 \begin{equation}
\langle Q_f,t_f|Q_i,t_i\rangle_{QPI}=\int \mathcal{D}''Q\mathcal{D}P\,\exp\left[i\int_{t_0}^t\,i\,dt' d\theta d\bar\theta\,\circled{$-i\cfrac{\bar \theta \theta}{\hbar}$}\,L[\Phi]\right]
\label{eq:29-2}.
 \end{equation}
We have encircled the quantities \circled{$\mathbb I$} and \circled{$-i\cfrac{\bar \theta \theta}{\hbar}$} because they seems to be the only quantities which are different in QM and CM. They somehow modify the {\it measure of integration} over the superspace $\int dtd\theta d\bar\theta$. We can extend the formalism also to the generating functionals as we have indicated in the introduction. 
\section{General Strategy\label{Section:3}}
The presence of a factor in the measure, both in Eq.~(\ref{eq:29-1}) and Eq.~(\ref{eq:29-2}), is reminiscent of another factor which appears in the measure of integration. This happens in Riemannian geometry. There we have distances defined via a metric $g_{\mu\nu}$ as 
\[
ds^2=g_{\mu\nu}\,dx^{\mu}\,dx^{\nu}
\]  
and we require that this distance is invariant under general coordinate transformations 
\[
x'^{\mu}=x'^{\mu}(x^{\nu}).
\]
We also require that the volume of integration is invariant and this happens only if we multiply the volume by a factor $E$:
\begin{equation}
\int \prod_{\nu=1}^4 \,dx^v\,E
\label{eq:31-1}.
\end{equation}
The factor $E$ is a determinant which is built in this way.  Let us called a tensor called vierbein $e^a_{\,\mu}$  which carries an index $a$ transforming under Lorentz transformations and a second index $\mu$ transforming under general coordinate transformations. It is possible to show that the metric $g_{\mu\nu}$ can be written in terms of the vierbein as follows 
\begin{equation}
g_{\mu\nu}=\eta_{ab}\,e^a_{\,\mu}\,e^b_{\,\nu},
\label{eq:32-1}
\end{equation} 
where $\eta_{ab}$ is a flat-Lorentz metric. For a review the interested reader can look into \citep{REF9}. The factor $E$ making the measure invariant is defined as 
\begin{equation}
E=\textrm{det} \,e^a_{\,\mu}
\label{eq:32-2}.
\end{equation}
In our case the space on which we would like to  introduce the factor $E$ is not the 4-dim. space time but the 3-dim. space $z^A=(t,\theta,\bar\theta)$.
Riemannian spaces with grassmannian coordinates have been studied in \citep{REF10}. We can define flat supertime in many ways but we choose the following one:
\begin{equation}
dz^A\,\eta_{AB}\,dz^B=dt^2-d\theta d\bar\theta+d\bar\theta d\theta,
\label{eq:33-0}
\end{equation}
where $\eta_{AB}$ is 
\begin{equation}
\eta_{AB}=\begin{pmatrix}
1&0&\,0\\
0&0&-1\\
0&1&\,0\\
\end{pmatrix}.
\label{eq:33-1}
\end{equation}
The analog of the Lorentz transformation in this case is given by the group $Osp(1|2)$, which is the set of transformations leaving invariant the quantity 
\begin{equation}
s=t^2+\theta\bar\theta-\bar\theta \theta
\label{eq:33-2}.
\end{equation}
The {\it non-flat} infinitesimal distance is defined as 
\begin{equation}
dz^A\,g_{AB}\,dz^B,
\label{eq:34-1}
\end{equation}
where $g_{AB}$ is the analog of the metric, but due to the grassmannian character of some of its elements, is called supermetric. Under a general superdiffeomorphisms of our coordinates, which we will indicate as: 
\begin{equation}
z^A=z^{A'} +\xi^A (z),
\label{eq:34-2}
\end{equation} 
the $g_{AB}$, in order for Eq.~(\ref{eq:34-1}) to be invariant, must transform as \citep{REF10}:
\begin{equation}
g_{AB}'=g_{AB}(z)+\frac{\overset{\rightarrow}{\partial}\xi^C}{\partial z^A}\, g_{CB}+g_{AC}\,\frac{\xi^C\overset{\leftarrow}{\partial}}{\partial z^B}+g_{AB,C}\,\xi^C,
\label{eq:34-3}
\end{equation}
where the right and left derivatives above are due to the grassmannian character of some of the $z$ and the fact that in Eq.~(\ref{eq:34-1}) some infinitesimals are to the left and some to the right. 
Like in normal Riemannian geometry also in super-Riemannian one \citep{REF10} we can define the super-vierbein which we will indicate with $E^A_{\Lambda}$, where $A$ is the Lorentz analog (Osp(1,2)) index and $\Lambda$ the general covariant (in supertime) one. The relation between supermetric and super-vierbein is \citep{REF10}:
\begin{equation}
g_{\Lambda\Pi}=E^{A}_{\,\,\Lambda}\,\eta_{AB}\,(-1)^{(1+B)\Pi}E^B_{\,\,\Pi}(z).
\label{eq:35-1}
\end{equation}
The numbers which are in the exponent of $(-1)$ are $0$ for $t$ and $1$ for $\theta$ and $\bar \theta$. For more details about grassmannian number, matrices and super-determinant (which are often indicated by $\textrm{sdet}(\dots)$) the reader can consult \citep{REF6} or the  \ref{Appendix:A} of this paper.
The analog of Eq.~(\ref{eq:31-1}) for the superspace made of $t,\theta,\bar\theta$ will be 
\begin{equation}
\int i\,dt \,d\theta\,d\bar\theta \,E
\label{eq:36-1}
\end{equation}
where $E=\textrm{sdet}(E^{A}_{\,M})$. If we compare this with Eq.~(\ref{eq:29-1}) and Eq.~(\ref{eq:29-2}) we can say that the CPI is like a   
``gauge'' fixed version of a ``super-general covariant'' formalism in supertime  and the ``gauge fixing'' is such that 
\begin{equation}
E=\mathbb I
\label{eq:36-2}
\end{equation}
while for the QPI the ``gauge fixing'' is such that 
\begin{equation}
E=-i\frac{\bar\theta\theta}{\hbar}.
\label{eq:37-1}
\end{equation}
Before going on further we should remember that Eq.~(\ref{eq:36-2}) and Eq.~(\ref{eq:37-1}) are not the only conditions we have to impose in order to obtain respectively the CPI and the QPI. We should infact remember that in a ``general covariant'' formalism also in the kinetic piece of the Lagrangian there is the presence of the vierbeins. Let us first suppose we integrate out  in Eq.~(\ref{eq:29-1}) and Eq.~(\ref{eq:29-2}) the $P$ so that the kinetic piece in both of them is reduced to 
\begin{equation}
\partial_t Q\partial_t Q,
\label{eq:37-2}
\end{equation}
where we omit the indices on $Q$. An analog ``general covariant'' piece would be 
\begin{equation}
D_tQ\,D_tQ
\label{eq:38-1}
\end{equation}
where the general covariant derivative $D_t$ would be 
\begin{equation}
D_t=E^M_{\,\,\,t}\partial_M,
\label{eq:38-2}
\end{equation}
with $E^M_{\,\,\,t}$ components of the {\it inverse} of the vierbein matrix appearing in Eq.~(\ref{eq:35-1}) and Eq.~(\ref{eq:36-1}). 
The $E^M_t$ should be chosen to be real because expression Eq.~(\ref{eq:38-1}) is real. If instead Eq.~(\ref{eq:38-1}) were of the form 
\[
(D_tQ)(D_tQ)^*,
\]
then we could choose $E^M_t$ to be complex. We will extend the reality condition of the $E^M_t$ to all the components of the vierbein in order to simplify the treatment. 

For the expression Eq.~(\ref{eq:38-1}) to be the same  as Eq.~(\ref{eq:37-2}), we will see later on that we have to make a particular choice for the vierbein. This choice, beside the Eq.~(\ref{eq:36-2}) for the CPI  and the Eq.~(\ref{eq:37-1}) for the QPI, is something like a gauge fixing that we need to impose on the "general covariant" formalism where the vierbein are free.  
The reader may  object that we should also insert a Faddev-Popov (F.P.) determinant in the functional measure. As we already said earlier, we think this is not necessary in our two cases because the F.P. would depend only on $E^A_{\,M}$ in our two gauge-fixings  and  we do not have the integration over $E^A_{\,M}$ in the path-integral.  Of course what we get is not a ``gauge fixing'' independent formalism neither one in which we can pass from the CPI to the QPI via a ``gauge transformation''. So we should be careful in saying that the CPI and the QPI are something like a ``gauge fixing'' of a general covariant formalism. In fact we have
used the expression ``something like''. Nevertheless we think that is worth to pursue this analogy and see if it helps us better undertand the interplay between CM and QM
\section{Vierbeins\label{Section:4}}
In this section we shall build the vierbein $E^M_{\,A}$, which gives the CPI and the one which gives the QPI. We will bother the reader with several details which are crucial in order to get the precise form of vierbeins.
We will show that there is a whole family of $E^{A}_{\,B}$, which reproduce the same CPI and the same for the QPI. The vierbein is a $3\times 3$ super matrix 
\begin{equation}
E^M_{\,A}=\begin{pmatrix}
a&\alpha&\beta\\
\gamma&b&c\\
\delta&d&e\\
\end{pmatrix},
\label{eq:41-1}
\end{equation} 
where the greek letter indicate an odd element while the latin one indicates an even element. It is easy to see why the elements of $E^A_{\,M}$ have the features indicated above by considering how the supermetric $g_{AB}$ is built out of the vierbein Eq.~(\ref{eq:35-1}) and the odd/even characters of the elements  of the $g_{AB}$. The two conditions that we have to satisfy to get the CPI are:  
\begin{equation}
\left\{
\begin{matrix}
E=1 \Longrightarrow \textrm{sdet}(E^M_{\,A})=1 \\
\\
D_tQ\,D_fQ=\partial_t Q\,\partial_t Q\quad.
\end{matrix}
\right.
\label{eq:42-1}
\end{equation}
It is a long calculation, reported in \ref{Appendix:B}, to prove that the vierbein for the CPI, satisfying the constraints (\ref{eq:42-1}) is given by 
\begin{equation*}
E^M_{\,A}(CPI)=\begin{pmatrix}
\pm 1&0&0\\
\gamma&b&c\\
\delta&d&e\\
\end{pmatrix}.
\end{equation*} 
where the variables $b,c,d,e$ have to satisfy two constraints reported in  \ref{Appendix:B}. A similar but much longer calculation, reported in  \ref{Appendix:C}, gives the form of the vierbein for the QPI 
\begin{equation*}
E^M_A(QPI)=\begin{pmatrix}
 1+a_S \theta\bar\theta& \alpha&\beta\\
\gamma&b&c\\
\delta&d&e 
\end{pmatrix}, 
\end{equation*} 
where $a_S$ is the \invcomma{soul} (see  \ref{Appendix:A} or ref.\citep{REF6} for the definition of soul). Also the elements of $E^M_A(QPI)$ are not free but must satisfy two constraints presented in \ref{Appendix:A}. Of course for the QPI also Eq.~(\ref{eq:42-1}) is different and it is reported in details in  \ref{Appendix:C}.
 %%%%%%%%%%%%%%%%%%%%%%%%%%%%%%%%%%%%%%%%%%%%%%%%
%%%%%%%%%%%%%%%%%%%%%%%%%%%%%%%%%%%%%%%%%%%%%%%%
%%%%%%%%%%%%%%%%%%%%%%%%%%%%%%%%%%%%%%%%%%%%%%%%
\section{Metrics \label{Section:5}}
In this section we will calculate the metric from the vierbeins and show that they depend on a lower number of free parameters than the vierbeins. Also in this section we will bother the reader with details but they are crucial in order to build the metric with the least number of parameters. We will skip the similar details in later sections for the curvatures because most of those calculations were done using {\it Wolfram Mathematica} and using symbols already defined in this and the previous section.\\ 
Let us start from the CPI. The vierbein, before implementing the constraint Eq.~(\ref{eq:43-2}), has the form Eq.~(\ref{eq:43-1}), i.e.
\begin{equation}
E^M_{\,A}=\begin{pmatrix}
\pm 1&0&0\\
\gamma&b&c\\
\delta&d&e\\
\end{pmatrix},
\label{eq:58-1}
\end{equation} 
and the \invcomma{super-metric} has the following form as a function of the vierbein: 
\begin{equation}
g^{\Lambda\Pi}=E^{\Lambda}_{\,\,A}\,\eta^{AB}\,(-1)^{(1+B)\Pi}E^{\Pi}_{\,\,B}.
\label{eq:58-2}
\end{equation}
A long but easy calculation gives 
\begin{equation}
g^{MN}=\begin{pmatrix}
1-2\,\gamma\,\delta &\gamma\,d-\delta\,b&\gamma\,e-\delta\,c\\
\gamma\,d-\delta\,b &0&b\,e-c\,d\\
\gamma\,e-\delta\,c&-(b\,e-c\,d)&0
\end{pmatrix}
\label{eq:59-1}
\end{equation}
and implementing the constraint (\ref{eq:43-2}) we get: 
\begin{equation}
g^{MN}=\begin{pmatrix}
1-2\,\gamma\,\delta &\gamma\,d-\delta\,b&\gamma\,e-\delta\,c\\
\gamma\,d-\delta\,b &0&\pm 1\\
\gamma\,e-\delta\,c&\mp 1&0
\end{pmatrix}.
\label{eq:59-2}
\end{equation}
Apparently this metric depends on $\gamma,\delta,b,c,d,e$ which means 12 parameters minus the two constraints on $b,c,d,e$ so on 10 parameters.
Actually the combinations of parameters which enter the $g^{MN}$ is less. In fact let us define the following variables:
\begin{equation}
\left\{
\begin{matrix}
\pi_1&\equiv&\gamma_{\theta}\,e_B-\delta_{\theta}\,c_B\\
\pi_2&\equiv&\gamma_{\bar\theta}\,e_B-\delta_{\bar\theta}\,c_B\\
\pi_3&\equiv&\delta_{\theta}\,b_B-\gamma_{\theta}\,d_B\\
\pi_4&\equiv&\delta_{\bar\theta}\,b_B-\delta_{\bar\theta}\,d_B
\end{matrix}
\right.
\label{eq:60-1}
\end{equation}
and 
\begin{equation}
\pi_5\equiv \gamma_{\bar\theta}\,\delta_{\theta}-\gamma_{\theta}\,\delta_{\bar\theta}. 
\label{eq:60-2}
\end{equation}
This $\pi_5$ is actually dependent on the other four $\pi_i$ of Eq.~(\ref{eq:60-1}), in fact it is easy to show that 
\[
\pi_5=\pm(\pi_2\,\pi_3-\pi_1\,\pi_4).
\]
It is easy to see that the metric Eq.~(\ref{eq:59-1}) can be written in term of the $\pi_i$ as follows
\begin{equation}
g^{MN}=\begin{pmatrix}
1\pm 2\bar\theta\theta(\pi_2\pi_3-\pi_1\pi_4)&-\pi_3\theta-\pi_4\bar\theta&\pi_1\theta+\pi_2\,\bar\theta\\
-\pi_3\theta-\pi_4\bar\theta&0&\pm 1&\\
\pi_1\theta+\pi_2\,\bar\theta&\mp 1&0
\end{pmatrix}.
\label{eq:60-3}
\end{equation}
Later on, in order to build the Christoffel symbols and the various curvatures tensor, we shall need also the inverse of $g^{MN}$ which turns out to have the following expression: 
\begin{equation}
g_{MN}=\begin{pmatrix}
1&\mp(\theta\pi_1+\bar\theta \pi_2)&\mp(\theta\pi_3+\bar\theta \pi_4)\\
\pm(\theta\pi_1+\bar\theta \pi_2)&0&\mp(1+\bar\theta\theta(\pi_2\,\pi_3-\pi_1\,\pi_4))\\
\pm(\theta\pi_3+\bar\theta \pi_4)&\pm(1+\bar\theta\theta(\pi_2\,\pi_3-\pi_1\,\pi_4))&0
\end{pmatrix}.
\label{eq:60-4}
\end{equation}
In both metrics above we have made the choice $a=\pm 1$ which is consistent with the CPI.
The reader may wonder why the metric has less free parameters than the vierbein. We feel the reason is because of the particular combination of vierbeins which enters the metric (see Eq.~(\ref{eq:58-1})). Moreover the vierbein is a more general object than the metric; in fact it enters the dynamics of particles of any spin. One last question the reader may have is if the $\pi_i$ are really free parameters or not. We feel they are free because they are made (see Eq.~(\ref{eq:60-1})) of combinations of odd variables $\gamma,\delta$ and even one $b,c,d,e$ and only these last ones are constrained (see Eq.~(\ref{eq:43-2}), while the first ones are totally free.  Let us now build the metric for the quantum case (QPI) or better for the \invcomma{regularized} quantum case. 
The very long details of the calculations are confined in \ref{Appendix:D}. The result is anyhow the following 

\begin{equation*}
g^{MN}=
\begin{pmatrix}
1-2\,\pi_5^Q\,\bar\theta\theta&-\pi_3^Q\,\theta-\pi_4^Q\,\bar\theta&\pi_1^Q\,\theta+\pi_2^Q\,\bar\theta\\
-\pi_3^Q\,\theta-\pi_4^Q\,\bar\theta&0&\pi_7^Q+\pi^Q_6\bar\theta\theta\\
\pi_1^Q\theta+\pi_2^Q\bar\theta&-\pi_7^Q-\pi_6^Q\,\bar\theta\theta&0
\end{pmatrix}.
%  \label{eq:68-1},
\end{equation*}
where the variables $\pi_1^Q,\dots, \pi_7^Q$ are properly defined in \ref{Appendix:D}.
 %%%%%%%%%%%%%%%%%%%%%%%%%%%%%%%%%%%%%%%%%%%%%%%%
%%%%%%%%%%%%%%%%%%%%%%%%%%%%%%%%%%%%%%%%%%%%%%%%
%%%%%%%%%%%%%%%%%%%%%%%%%%%%%%%%%%%%%%%%%%%%%%%%
\section{Curvatures \label{Section:6}}
In this section we will build the curvatures from the metric presented in the previous section. As the calculations are very long we have made used of a package of {\it Wolfram Mathematica } dedicated to calculations containing grassmannian variables \citep{REF14}. The same package has been used also for calculating the metric of the previous section and other calculations presented all through the paper. The first thing we have calculated has been the Christoffel symbols associated to our varoius metrics. If we work in a space with odd and even variables the Christoffel symbols $\Gamma^C_{AB}$ have the following expression \citep{REF10}:
\begin{eqnarray}
\Gamma_{AB}^C&=&(-)^{BC}\frac{1}{2}\left[(-)^{BD}\,g_{AD,B}+(-)^{A+B+AB+AD}\,g_{BD,A}-g_{AB,D}\right]\,g^{DC}\nonumber\\
 \label{eq:77-1}
\end{eqnarray}
where the comma on the metric like $g_{AD,B}$ means the derivative of $g_{AD}$ respect to the variable $B$ . As usual the exponent on the $(-)$ indicate the even $(0)$ or odd $(1)$ nature of the associated variables. The results of the calculations of  Eq.~(\ref{eq:77-1}) for both the CPI metric Eq.~(\ref{eq:60-3}) and  Eq.~(\ref{eq:60-4}) and for the QPI one Eq.~(\ref{eq:68-3}) and Eq.~(\ref{eq:75-1}) are confined in \ref{Appendix:E}.
Once we have the Christoffel symbols we can calculate the various curvatures. The definition we will use is the one of  ref.\citep{REF10} for SuperRiemannian space:
\begin{eqnarray}
R_{ABC}^D&=&-\Gamma_{AC,B}^D+(-)^{BC}\,\Gamma_{AB,C}^D-(-)^{C(D+E)}\,\Gamma_{AC}^E\,\Gamma_{EB}^D+\nonumber\\
&(-)&^{B(C+D+E)}\,\Gamma_{AB}^E\,\Gamma_{EC}^D.
 \label{eq:78-1}
\end{eqnarray}
From this we can build the Ricci curvature tensor defined as 
\begin{equation}
R_{AB}\equiv (-)^C\,R_{ABC}^C.
 \label{eq:78-2}
\end{equation}
Its expression in term of Christoffel symbols is: 
\begin{eqnarray}
R_{AB}&=&(-)^{C+1}\,\Gamma_{AC,B}^C+(-)^{C(B+1)}\,\Gamma_{AB,C}^C+\\
&-&(-)^{C(C+E-1)}\,\Gamma_{AC}^E\,\Gamma_{EB}^C+(-)^{BE+C}\,\Gamma_{AB}^E\,\Gamma_{EC}^C.\nonumber
 \label{eq:78-4}
\end{eqnarray}
From the Ricci curvature tensor we can calculate the so-called Ricci scalar defined in \citep{REF10} as 
\begin{equation}
R=(-)^{B}\,g^{BA}\,R_{AB}.
 \label{eq:79-1}
\end{equation}
The explicit expression of all components of the Ricci tensor and of the Ricci scalar has been confined to \ref{Appendix:F} and \ref{Appendix:G}. Its calculation, again, has been made possible by the use of {\it Wolfram Mathematica} \citep{REF14}.
The things to notice for the curvatures of the QPI (see Eq.~(\ref{eq:C-3-1}) and (\ref{eq:D-4-3})) is they are singular for $\epsilon\to 0$. Infact in many of its component we have the $\pi^Q_7$ in the denominator and $\pi^Q_7$ is proportional to $\epsilon$. 
Only the $R_{\theta\theta}$ and $R_{\bar\theta\bar\theta}$ are not singular because they are equal to zero. What we should take care of  is the Ricci
  scalar,  Eq.~(\ref{eq:C-4-1}) and (\ref{eq:D-4-4})), where the singularity cannot be a coordinate artifact because it is a scalar independent of the
  coordinates. The way out could came from the fact that in the true quantum case ($\epsilon\to 0$) se also have that $\theta$ and $\bar\theta$
  have to be sent to zero \citep{REF10}. So for example in Eq.~(\ref{eq:C-4-1})  we have that the fourth  contribution is proportional to $\bar\theta\theta/\pi_7$ that would give a $0/0$, which is an undefined term. But being $\epsilon$ and $\bar\theta\theta$ totally independent, we could choose that this undefined form is a finite grassman number. The next term is proportional to $\theta\bar\theta$ and it would go to zero. In this manner the Ricci scalar would not blow up in QM, which seems a natural thing to require as nothing goes to infinity in QM 
\section{Zeros of the curvature\label{Section:7}}
In this section we will check if there are value of the $\pi_i$ for which the various curvatures turn out to be zero or at least some of them.  
As the $\pi_i$ are {\it arbitrary} we can choose the system to sit on those values and so conclude that those curvatures are zero. This is what happens in the CPI as we will check.  Surprisingly this does not happens in the QPI. There is no point where the curvatures vanish. This \invcomma{may} indicate that there is some sort of \invcomma{\underline{hidden} \underline{matter}} in QM. Of course we don't identify this with the so called \invcomma{hidden variables} of Einstein \citep{REF13}.
\subsection{Zeros of the curvature in the CPI}
 Let us start with the CPI in the case where the $\pi_i$ are indipendent of time. The Ricci scalar was given in Eq.~(\ref{eq:C-2-1}) and it had the following expression: 
\begin{eqnarray}
R^{CPI}&=&-\frac{1}{2}\left(\pi_2^2-22\,\pi_2\,\pi_3+\pi_3^2+20\,\pi_1\,\pi_4\right)+\nonumber\\
&+&8\,\frac{\bar\theta\theta}{a}\,\left(\pi_2\,\pi_3-\pi_1\,\pi_4\right)^2.
\label{eq:82-1}
\end{eqnarray}
In order to have $R^{CPI}=0$ we need to have zero both its soul and body, i.e.:
\begin{equation}
\label{eq:83-1}
\begin{lcases}
&\pi_2\,\pi_3-\pi_1\,\pi_4=0\\
&\pi_2^2-22\,\pi_2\,\pi_3+\pi_3^2+20\,\pi_1\,\pi_4=0.
\end{lcases}
\end{equation}
From the first equation in  (\ref{eq:83-1}) we get 
\begin{equation}
\label{eq:83-2}
\pi_1=\frac{\pi_2\,\pi_3}{\pi_4}
\end{equation}
and putting this into the second of Eq.~(\ref{eq:83-1}) we get 
\begin{equation}
\label{eq:83-3}
\pi_2=\pi_3.
\end{equation}
In the space described by the four parameters $\pi_1,\pi_2,\pi_3,\pi_4$ the Ricci scalar is zero on a 2-dim surface descibed by 
Eq.~(\ref{eq:83-2}) and Eq.~(\ref{eq:83-3}). Let us now check if on this surface, or at least on some points, also the Ricci tensor is zero. Let us look at the various components of the Ricci tensor presented in Eq.~(\ref{eq:C-1-1}). Let us start with 
\[
R_{tt}=\frac{1}{2}\left(\pi_2+\pi_3\right)^2-2\,\pi_1\,\pi_4.
\]   
Using Eq.~(\ref{eq:83-2}) and Eq.~(\ref{eq:83-3}) it is straightforward to show that $R_{tt}=0$.
Let us now move on to 
\begin{equation*}
R_{t\theta}=-R_{\theta t}=\frac{(\theta\,\pi_1+\bar\theta\,\pi_2)\,(-(\pi_2+\pi_3)^2+4\,\pi_1\,\pi_4)}{2\,a}.
\end{equation*}
The second factor on the right is zero on Eq.~(\ref{eq:83-1}) so $R_{t\theta}=R_{\theta t}=0$. Let us now check 
\[
R_{t\bar\theta}=-R_{\bar\theta t}=-\frac{(\theta\,\pi_3+\bar\theta\,\pi_4)\,((\pi_2+\pi_3)^2-4\,\pi_1\,\pi_4)}{2\,a}.
\]   
Again the second factor on the right is zero on Eq.~(\ref{eq:83-1}) so  $R_{t\bar\theta}=R_{\bar\theta t}=0$. Next let us check 
\begin{eqnarray*}
R_{\theta\bar\theta}&=&\frac{1}{2}\left[\bar\theta\theta \underbrace{\left(\pi_1\,\pi_4-\pi_2\,\pi_3\right)}_A\,\left(\pi_2^2-6\,\pi_2\,\pi_3+\pi^2_3+4\,\pi_1\,\pi_4\right)\right.\nonumber\\
&-&a\,\left.\underbrace{\left(\pi_2^2-10\,\pi_2\,\pi_3+\pi^2_3+8\,\pi_1\,\pi_4\right)}_B\right].
\end{eqnarray*}
 %%%%%%%%%%%%%%%%%%%%%%%%%%%%%%%%%%%%%%%%%%%%%%%%
%%%%%%%%%%%%%%%%%%%%%%%%%%%%%%%%%%%%%%%%%%%%%%%%
%%%%%%%%%%%%%%%%%%%%%%%%%%%%%%%%%%%%%%%%%%%%%%%%
The term $A=0$ because of Eq.~(\ref{eq:83-1}) while $B$, using Eq.~(\ref{eq:83-1}), can be transformed as follows: 
\begin{equation*}
B=\pi_2^2-10\,\pi_2\,\pi_3+\pi^2_3+8\,\pi_1\,\pi_4=\pi_2^2-10\,\pi_2^2+\pi^2_2+8\,\pi_2^2=0.
\end{equation*}
So $R_{\theta\bar\theta}=-R_{\bar\theta\theta}=0$. \\
The reason why we can choose the values of the $\pi_i$ on which our curvature is zero is because the  $\pi_i$ do not enter the lagrangian of the CPI and we can change them without the Lagrangian getting modified. So far in the CPI we have proved that both the Ricci scalar and the Ricci curvature can be brought to zero. This is a situation very similar to the Schwarzschild case where, for points outside the mass region we have both $R$ and $R_{ab}$ equal to zero. What is not zero there is another scalar built up from the curvature tensor:
\[
R_{abcd}R^{abcd}\neq 0. 
\]
This quantity in the Schwarzschild case is proportional to $G/r^6$ and it is the indicator of the presence of matter somewhere. We should calculate the analog quantity for the CPI. In this case the quantity to calculate is:
\[
R_{abc}R^{abc}. 
\]
Instead of doing this rather complicated calculation, we should remind ourselves that our analog of space-time is $(t,\theta,\bar\theta)$ so we should just check if our Ricci scalar and tensor are zero for any value of $(t,\theta,\bar\theta)$. This would not happen in the Schwarzschild case in the area where there is matter. The calculation we have done is without $t$ and with $\theta,\bar\theta\neq 0$ and it gives zero everywhere. Let us now generalize it to the case with $t$ present that means when the $\pi_i$ depend on $t$ and see if we get zero everywhere in $(t,\theta,\bar\theta)$. 
Let us start by finding the points on which the Ricci scalar is zero for $\pi_i$ depending on time. The $R^{CPI}$ is given by Eq.~(\ref{eq:D-4-2}) 
\begin{eqnarray}
R^{CPI}&=&-\frac{1}{2}\left[\pi_2^2-22\,\pi_2\pi_3+\pi^2_3+20\,\pi_1\pi_4+4(\pi_3'-\pi_2 ')\right]\\
&+&\frac{\bar\theta\theta}{a}\left[8\,(\pi_2\pi_3-\pi_1\pi_4)^2+4\,a\,(\pi_3'-\pi_2')\,\pi_5+7\,a\,\pi_5'(\pi_3-\pi_2)+2\,a\,\pi_5''\right].\nonumber
\end{eqnarray}
Both the soul and the body must be zero, i.e.
\begin{equation}
\label{eq:88-1}
\begin{lcases}
&\pi_2^2-22\,\pi_2\pi_3+\pi^2_3+20\,\pi_1\pi_4+4(\pi_3'-\pi_2 ')=0\\
&8\,(\pi_2\pi_3-\pi_1\pi_4)^2+4\,a\,(\pi_3'-\pi_2')\,\pi_5++7\,a\,\pi_5'(\pi_3-\pi_2)+2\,a\,\pi_5''=0.
\end{lcases}
\end{equation}
If in the time-independent case Eq.~(\ref{eq:83-2}) and Eq.~(\ref{eq:83-3}) are choosen, starting from the relation between $\pi_5$ and $\pi_1,\pi_2,\pi_3,\pi_4$
\[
\pi_5=\frac{\pi_2\,\pi_3-\pi_1\,\pi_4}{a},
\] 
it can be easily seen that 
\begin{equation}
\label{eq:88-2}
\pi_5=0, 
\end{equation}
from which it follows that $\pi_5'=\pi_5''=0$. 
The  system of  Eq.~(\ref{eq:88-1}) consequently reduces to 
\begin{equation}
\begin{lcases}
&\pi_2^2-22\,\pi_2\pi_3+\pi^2_3+20\,\pi_1\pi_4=0\\
&8\,(\pi_2\pi_3-\pi_1\pi_4)^2=0.
\end{lcases}
\end{equation}
which is equivalent to the system of Eq.~(\ref{eq:83-1}), whose solutions Eq.~(\ref{eq:83-2}) and Eq.~(\ref{eq:83-3}) are the one we started with in the time independent case. It follows that the constraints on which $R^{CPI}$ is zero, even in the time dependent case, are  Eq.~(\ref{eq:83-2}) and Eq.~(\ref{eq:83-3}) like in the case of $\pi_i$ independent on time. Of course there may be other set of points on which it is zero (because $\pi_2=\pi_3$ was our choice) but what is important is that we have found points in which it is zero. Next we should check that also the Ricci tensor is zero on the same set of points.   
Their expression is given in (\ref{eq:D-4-1}) of \ref{Appendix:D} and it is easy to check they are all zero on the points where the following condition are satisfied (\ref{eq:83-2}),(\ref{eq:83-3}),(\ref{eq:88-2}): 
\begin{equation}
\label{Eq:50}
\begin{lcases}
&\pi_2=\pi_3\\
&\pi_1=\frac{\pi_2\,\pi_3}{\pi_4}\\
&\pi_5=0.
\end{lcases}
\end{equation}
Let us consider for example $R_{tt}$ of (\ref{eq:D-4-1}):   
\begin{align}
R_{tt}&=R_{tt}^{CPI}(\pi_i)+(\pi_3'-\pi_2')+\bar\theta\theta\left[\pi_3^2\pi_2'-\pi_2^2\pi_3'+(\pi_2-\pi_3)(\pi_4\pi_1'+\pi_4'\pi_1)+\right.\nonumber\\
&+\left.(\pi_3'-\pi_2')(a\,\pi_5-2\,\pi_2\pi_3+3\,\pi_1\pi_4)+2\,a\pi_5'\right].
\end{align}
 It results that the first term is equal to zero because it has the same expression as the time independent $R_{tt}^{CPI}$, which was zero on the contraints (\ref{eq:83-2}),(\ref{eq:83-3}). As for the other contributions they are trivially zero because the conditions $\pi_2=\pi_3$ and $\pi_5=0$ imply $\pi_2'=\pi_3'$ and $\pi_5'=0$. In the same way it can be shown that all the other contributions are identically equal to zero. 
 As we explainied in the time independent case we do not calculate 
\[
R_{abc}R^{abc}, 
\]
bacause we proved that the Ricci scalar and the tensors are zero over the whole $(t,\theta,\bar\theta)$ space. As a consequence we have that there is no matter anywhere differently than in the Schwarzschild case. 
\subsection{Lack of zeros in the QPI curvature}
Let us now turn to the QPI case starting with the Ricci scalar in the case independent on time. Its expression was given in Eq.~(\ref{eq:C-4-1}):
\begin{eqnarray}
\label{eq:92-1}
R^{QPI}&=&R^{CPI}(\pi^Q_i)+(2\,\sigma_1-3\,\pi^Q_6)+\\
 &+&\frac{\bar\theta\theta}{\pi_7^Q}\left[-(\pi_2^{Q^2}+6\,\pi_2^Q\pi_3^Q+\pi_3^{Q^2}-8\,\pi_1^Q\pi_4^Q)\pi_6^Q+4\,\pi_6^{Q^2}+\right.\nonumber\\
 &-&\left.4\,\sigma_1(\pi_2^{Q^2}+3\,\pi_2^Q\pi_3^Q+\pi_3^{Q^2}-5\,\pi_1^Q\pi_4^Q-\pi_6^Q+\sigma_1)\right].\nonumber
\end{eqnarray}
The $\sigma_1$ was defined in (\ref{eq:B-1-1bis}) of \ref{Appendix:E} and the parameters of (\ref{eq:92-1}) are not four but six. The various $\pi_i^Q$ were introduced in (\ref{eq:70-1}). Note that $\pi_5^Q$ does not appear because it is related to the others. Let us suppose we stay on the surface where
\begin{equation}
\label{eq:92-2}
R^{CPI}(\pi_i^Q)=0.
\end{equation}
and this happens when the constraints Eq.~(\ref{eq:83-2}) and Eq.~(\ref{eq:83-3}) are satisfied but with the $\pi_i$ replaced by the $\pi_i^Q$.
Next, for the body of Eq.~(\ref{eq:92-1}) to be zero we need that 
\[
2\,\sigma_1-3\,\pi^Q_6=0
\]
i.e.
\begin{equation}
\sigma_1=\frac{3}{2}\,\pi_6^Q.
\label{eq:93-1}
\end{equation}
Like in the CPI, we keep $\pi_4^Q$ and $\pi_3^Q$ free and link the other vabiables to these two via the $R^{CPI}(\pi_i^Q)=0$.
Also  $\pi_6^Q$ seems to be free and  the same $\pi_7^Q$, which does not make its appearance in Eq.~(\ref{eq:92-1}) but was present in the formalism. So Eqs.~(\ref{eq:93-1}) and (\ref{eq:92-2}) make the body of $R^{QPI}$ zero;  now we have to make the soul zero. Using the constrain 
Eq.~(\ref{eq:92-2}), which leads to $\pi_2^Q=\pi_3^Q$, and Eq.~(\ref{eq:93-1}), after straightforward calculations we get that the soul of $R^{QPI}$  is given by
\[
\textrm{soul}(R^{QPI})=\frac{\pi^{{Q}^2}_6}{\pi_7^Q}.
\]
So to be zero we have to set 
\begin{equation}
\pi_6^Q=0.
\label{eq:94-1}
\end{equation}
 %%%%%%%%%%%%%%%%%%%%%%%%%%%%%%%%%%%%%%%%%%%%%%%%
%%%%%%%%%%%%%%%%%%%%%%%%%%%%%%%%%%%%%%%%%%%%%%%%
%%%%%%%%%%%%%%%%%%%%%%%%%%%%%%%%%%%%%%%%%%%%%%%%
We can summarize the set of constraints which make $R^{QPI}=0$ as:
\begin{equation}
\label{eq:94-2}
\begin{lcases}
\pi_2^Q&=\pi_3^Q\\
\pi_1^Q&=\cfrac{\pi_2^Q\,\pi_3^Q}{\pi_4^Q}\\
\pi_6^Q&=0\\
\sigma_1&=\cfrac{3}{2}\,\pi_6^Q
\end{lcases}
\end{equation}
Let us now see  if also the Ricci tensor for the QPI, given by  Eq.~(\ref{eq:C-3-1}) is zero on the points of Eq.~(\ref{eq:94-2}). Let us start from $R_{tt}^{QPI}$:
\begin{eqnarray*}
R_{tt}^{QPI}&=&\underbrace{R_{tt}^{CPI}(\pi_i^Q)}_{A}+2\,\sigma_1+\underbrace{\frac{\bar\theta\theta}{\pi_7^Q}\left[((\pi_2^Q+\pi_3^Q)^2-4\,\pi_1^Q\pi_4^Q)\pi_6^Q\right.}_{B}\nonumber\\
&+&\left.\underbrace{2\,(\pi_2^{Q^2}+4\,\pi_2^Q\pi_3^Q+\pi_3^{Q^2}-6\,\pi_1^Q\pi_6^Q-\pi_6^Q)\,\sigma_1+6\sigma_1^2}_{C}\right].
\end{eqnarray*}
The $A$-term calculated using Eq.~(\ref{eq:94-2}) gives  
\begin{eqnarray*}
A&=&\left[\frac{1}{2}(\pi_2^Q+\pi_3^Q)^2-2\,\pi_1^Q\pi_4^Q\right]\\
&=&\left[
\frac{1}{2}(2\,\pi_2^Q)^2-2\,\pi_2^Q\pi_3^Q
\right]=2\,\pi_2^{Q^2}-2\,\pi_2^{Q^2}=0.
\end{eqnarray*}
The term $B$ is zero because is multiplied by $\pi_6^Q$ which is zero by Eq.~(\ref{eq:94-2}). The term $C$ is zero because $\sigma_1=0$. After the $A$ piece there is a $\sigma_1$ which is zero. 
So $R_{tt}^{QPI}=0$. Let us now analyze the $R_{t\theta}^{QPI}$ which is 
 \begin{eqnarray}
\label{eq:96-1}
 R_{t\theta}^{QPI}&=&-R_{\theta t}^{QPI}\\&=&R_{t\theta}^{CPI}(\pi_i^Q)-\frac{\pi_6^Q+3\,\sigma_1}{\pi_7^Q}\left[\theta\pi_1^Q+\bar\theta(\pi_2^Q+\pi_3^Q)\right].\nonumber
 \end{eqnarray}
 The $R_{t\theta}^{CPI}=0$ on the constraints  Eq.~(\ref{eq:94-2}). The second piece in  Eq.~(\ref{eq:96-1}) is zero because $\pi_6^Q=\sigma_1=0$, so $R_{t\theta}^{QPI}=0$. Next let us analyze $R_{\theta\bar\theta}$ which is 
 \begin{eqnarray}
 R_{\theta\bar\theta}^{QPI}&=&-R_{\bar\theta\theta}=R_{\theta\bar\theta}^{CPI}(\pi_i^Q)+\frac{\sigma_1-3\,\pi_6^Q}{\pi_7^Q}\\
 &+&\frac{\bar\theta\theta}{\pi_7^{Q^2}}\left[2\,\pi_6^{Q^2}+8\,\pi_1^Q\pi_4^Q\pi_6^Q-8\,\pi_2^Q\pi_3^Q\pi_6^Q\sigma_1\left((\pi_2^Q-\pi_3^Q\right)^2+4\,\pi_6^Q+2\,\sigma_1)\right].\nonumber 
\label{eq:97-1}
 \end{eqnarray}
 Again $R_{\theta\bar\theta}^{CPI}=0$ is zero on the constraints Eq.~(\ref{eq:94-2}) and all the rest is zero because $\pi_6^Q=\sigma_1=0$. \par Now let us check if the constraint Eq.~(\ref{eq:94-2}) leads to any contradiction.
Let us go back to the definition of $\pi_6^Q$ given in Eq.~(\ref{eq:72-3})
\begin{equation*}
\pi_6^Q=\epsilon\,a_S-\cfrac{i}{\hbar}\,a_B\,(1-\epsilon)+a_B(\alpha_{\theta}\,\pi_2-\alpha_{\bar\theta}\,\pi_1+\beta_{\theta}\,\pi_4-\beta_{\bar\theta}\,\pi_3)+\alpha_{\bar\theta}\,\beta_{\theta}-\alpha_{\theta}\,\beta_{\bar\theta}.
\end{equation*}
Going to the true-quantum case $\epsilon=0$ we would get 
\begin{equation}
\label{eq:140bis}
\pi_6^Q=-\cfrac{i}{\hbar}\,a_B+a_B\underbrace{(\alpha_{\theta}\,\pi_2-\alpha_{\bar\theta}\,\pi_1+\beta_{\theta}\,\pi_4
-\beta_{\bar\theta}\,\pi_3)}_A+\underbrace{\alpha_{\bar\theta}\,\beta_{\theta}-\alpha_{\theta}\,\beta_{\bar\theta}}_B.
\end{equation}
 If, as we did before in our calculations, we choose $\alpha=\beta=0$, we get in the true quantum case:
\[
\pi_6^Q=-\cfrac{i}{\hbar}\,a_B
\]  
as $a_B=\pm 1$ we get  
\begin{equation}
\pi_6^Q=\mp\cfrac{i}{\hbar} 
\label{eq:98-1}\,\,.
\end{equation} 
 So it is never zero and this contradicts Eq.~(\ref{eq:94-2}) or, saying it better, because of Eq.~(\ref{eq:98-1}) the constraint Eq.~(\ref{eq:94-2})
 is not satisfied. The Ricci scalar curvature and the associated tensor are never zero in the true quantum case.
 Let us suppose we do not make the choice $\alpha=\beta=0$, then in Eq.~(\ref{eq:72-3}) we would have three terms: one which is a complex number 
 $\epsilon\,a_S-\cfrac{i}{\hbar}\,a_B\,(1-\epsilon)$ and the $A,B$ of Eq.~(\ref{eq:140bis}) which are the product of couple of grassmannian odd number like $\pi_2$ and $\alpha_{\theta}$ and similar. These $A$ and $B$ are grassmann even and they will never be equal to a complex number. So $A$ and $B$ cannot cancel the above mentioned complex number in order to put $\pi_6^Q$ equal to zero. As the $A$ and $B$ contain the parameterm $\pi_i^Q$ which are free they could  be put to zero, but then also the complex number has to be put to zero. This is possible in the regularized $QPI$ and it would mean   
 \[
\epsilon\,a_S-\frac{i}{\hbar}a_B(1-\epsilon)=0 
 \]
which is equivalent to: 
\begin{equation}
a_S=\frac{i}{\hbar}\,a_B\,\frac{1-\epsilon}{\epsilon}.
\label{eq:100-1}
\end{equation}
But in the true quantum case $\epsilon\to 0$ we would get $a_S\to \infty$, which does not make sense. Let us do the last attempt and see if we can put $R^{QPI}=0$ without putting $\pi_6^Q=0$. Let us go back to formula Eq.~(\ref{eq:92-1}) and let us see if there is a different method to get $R^{QPI}=0$.
If in Eq.~(\ref{eq:92-1}) we put first to zero the body and then the soul we get 
\[
R^{CPI}(\pi^Q_i)+2\sigma_1-3\pi_6^Q=0
\]
which leads to 
\begin{equation}
\pi_6^Q=\frac{1}{3}\,(R^{CPI}(\pi_i^Q)+2\sigma_1).
\label{eq:101-1}
\end{equation}
 %%%%%%%%%%%%%%%%%%%%%%%%%%%%%%%%%%%%%%%%%%%%%%%%
%%%%%%%%%%%%%%%%%%%%%%%%%%%%%%%%%%%%%%%%%%%%%%%%
%%%%%%%%%%%%%%%%%%%%%%%%%%%%%%%%%%%%%%%%%%%%%%%%
 %%%%%%%%%%%%%%%%%%%%%%%%%%%%%%%%%%%%%%%%%%%%%%%%
%%%%%%%%%%%%%%%%%%%%%%%%%%%%%%%%%%%%%%%%%%%%%%%%
%%%%%%%%%%%%%%%%%%%%%%%%%%%%%%%%%%%%%%%%%%%%%%%%
From this formula it seems that we do not have to put $\pi_6^Q=0$. But let us analize Eq.~(\ref{eq:101-1}) in detail. On the R.H.S. we have only terms which are product of grassmann variables like 
\[
\sigma_1=\pi_2^Q\pi_3^Q-\pi_1^Q\pi_4^Q-\pi_5^Q\pi_7^Q.
\]
The same for $R^{CPI}$ which is 
\begin{eqnarray*}
R^{CPI}(\pi_i^Q)&=&-\frac{1}{2}\left(\pi_2^{Q^2}-22\,\pi_2^Q\pi^Q_4+\pi_3^{Q^2}+20\,\pi_1^{Q^2}\pi_4^Q\right).\nonumber\\
&+&\bar\theta\theta\frac{8}{a}(\pi_2^Q\pi_3^Q-\pi_1^Q\pi_4^Q).
\end{eqnarray*}
So on the R.H.S. of Eq.~(\ref{eq:101-1}) we have products of 2 grassmannian numbers or higher terms (like those with $\bar\theta\theta$ which anyhow goes to zero in the true quantum case $\theta,\bar{\theta}\to 0$), while on the L.H.S. we have $\pi_6^Q$ which  (see Eq.~(\ref{eq:140bis})) contains both product of grassmannian numbers but also complex number which must be put to zero separately and we go back to the case descibed by Eq.~(\ref{eq:100-1}). So we can conlude that in the true quantum case we cannot bring the curvature to zero. 
 
\section{Conclusions\label{Section:8}}
In this paper we have shown that \invcomma{intrinsic} vierbein, persent in the CPI version of CM, gives zero curvature (at least the Ricci one). This seems natural because there is no \underline{external} \underline{mass} generating a curvature in the space on which our test particle of the CPI would move. In the quantum case there is an \underline{intrinsic} \underline{curvature}. One could immediately ask what the matter, which produce this curvature, is. We could speculate saying that there are some non-local hidden variables of the type Bell \citep{REF13} proposed long ago or it is some sort of dark matter or dark energy so fashionable these days. We do not know and we prefer not to speculate. Our goal at the beginning was to see if there was some \invcomma{intrinsic} geometry in Q.M and we feel we have found some hints of it. We also would like to notice that this intrinsic geometry appear when we look not only at the usual bosonic variables $\varphi^a$ of Q.M but also at their differential forms $c^a$, which we would like to call \invcomma{quantum forms}. This is a topic which has not been studied deeply except by few mathematicians and in a language difficult for physicists. We feel a more intense study should be done of this sector of mathematics. 
\newpage
\appendix
\section{\label{Appendix:A}}
\subsection{Grassmannian algebras}
Given a set of $N$-elements $\xi^a,a=1,\dots ,N$ obeying the following properties 
\[
\xi^a\xi^b=-\xi^b\xi^a,\,(\xi^a)^2=0,\textrm{for all $a,b$}, 
\]
they are called generators of a grassmannian algebra ${\Lambda}_N$.  \\The elements $1,\xi^a,\xi^a\xi^b,\xi^a\xi^b\xi^c,\dots$ form a set of $2^N$ objects, called the basis of the algebra.
An addition in this basis and a multiplication by complex numbers is defined among its elements and so they form a linear vector space of dimension $2^N$. 
\subsection{Super-numbers}
Every element $z$ of the vector space above can be written as 
\[
z=z_B+z_S
\]
where $z_B$ is an ordinary complex number and it is called the \invcomma{body} of $z$ and $z_S$, called the \invcomma{soul}, is:
\begin{equation}
z_S=\sum_{n=1}^{2^N}\frac{1}{n!}c_{a_1\cdots a_n}\xi^{a_n}\cdots \xi^{a_1},
\label{eq:A-2-1}
\end{equation} 
where the $c_{a_1\cdots a_n}$ are also complex numbers. The $c_{a_1\cdots a_n}$ are antisymmetric in the exchange of their indices.
It is easy to prove that 
\[
z_S^{N+1}=0.
\]
\subsection{Inverse of a super-number}
The inverse $z^{-1}$ of a super-number, defined by $z\,z^{-1}=1$, turns out to be 
\[
z^{-1}=z_B^{-1}\sum_{n=0}^{2^N} \left(-z_B^{-1}\,z_S\right),
\]
so if $z_B=0$ the inverse does not exist. 
\subsection{C-number and A-number}
Any super-number can be split into its even \invcomma{e} and odd \invcomma{o} part as 
\begin{eqnarray*}
z&=&e+o\\
e&\equiv& z_B+\sum_{n}\frac{1}{2n!}c_{a_2\cdots a_{2n}}\xi^{a_{2n}}\cdots \xi^{a_2}\\
o&\equiv& \sum_{n}\frac{1}{2n+1!}c_{a_1\cdots a_{2n+1}}\xi^{a_{2n+1}}\cdots \xi^{a_1}.
\end{eqnarray*}
If a super-number has only an \invcomma{$e$} part is called an even super-number while if it has only an \invcomma{$o$} part it is called an odd super-number.
The grassmann index of even or odd numbers is the number $2n$ or $2n+1$ modulo 2 so for even number is zero and for odd number is one.
Usually it is put as exponent  of $(-1)$ and is indicated with square brackets: $[e],[o]$.  
\subsection{Super-vectors and super-matrices}
Super-vectors are defined regorously in \citep{REF6}, but basically are rows or columns of super-numbers. The elements in the basis of these vectors are arranged in such a manner that the even elements \invcomma{$e$} come above the odd \invcomma{"o"} one, like
\begin{equation}
\begin{pmatrix}
e\\o
\end{pmatrix}
\label{eq:A-4-1}.
\end{equation}
If the basis has the form Eq.~(\ref{eq:A-4-1})
then a super-matrix $\mathcal K $ can always be arranged in the form  
\begin{equation}
K=\begin{pmatrix}
A&C\\D&B
\end{pmatrix},
\label{eq:A-4-2}
\end{equation}
where the elements of the super-matrices $A$ and $B$ are made of super-numbers while $C$ and $D$ are made of odd numbers.
More details are given in \citep{REF6}.
\subsection{Super-trace}
The super-trace of the matrix $\mathcal K$ is defined as 
\[
\textrm{str} \mathcal{K}=(-1)^{[i]}\mathcal{K}
_i^i,
\]
where $[i]$ is the grassmann index of the \invcomma{i} elements. 
\subsection{Super-determinant and its inverse}
For a standard matrix $X$ we know that the following relation holds between the variation \invcomma{$\delta$} of parameters entering the determinant and the ones entering the trace:
\[
\delta[\textrm{ln det}X]=\textrm{tr}[X^{-1}\delta X].  
\]
We use this relation to define the super-determinant in case of a super-matrix $\mathcal K$ which has the form $\mathcal K$ of Eq.~(\ref{eq:A-4-2}). The result \citep{REF6} is:
\begin{equation}
\textrm{sdet}\begin{pmatrix}
A&C\\D&B
\end{pmatrix}
=\textrm{det}(A-CB^{-1}D)(\textrm{det}B)^{-1}
\label{eq:A-6-1}
\end{equation}
where the symbol \invcomma{det} has the same meaning as if the entries were complex numbers.
It is also possible to define the inverse of the supermatrix $X$ as:
\begin{equation}
X^{-1}=\begin{pmatrix}
\widetilde{A}&\widetilde{C}\\
\widetilde{D}&\widetilde{B}
\end{pmatrix}
\label{eq:A-6VR-1}
\end{equation}
where
\begin{align*}
\widetilde{A}&=(\mathbb{I}-A^{-1}CB^{-1}D)^{-1}A^{-1}\\
\widetilde{C}&=-(\mathbb{I}-A^{-1}CB^{-1}D)^{-1}A^{-1}CB^{-1}\\
\widetilde{D}&=-(\mathbb{I}-B^{-1}DA^{-1}C)^{-1}B^{-1}DA^{-1}\\
\widetilde{B}&=(\mathbb{I}-B^{-1}DA^{-1}C)^{-1}B^{-1}.
\end{align*}
Note that this inverse exists if only $A$ and $B$ are not singular. It is also easy to calculate the determinant of the inverse:
\begin{equation*}
\textrm{sdet}\begin{pmatrix}
A&C\\D&B
\end{pmatrix}^{-1}
=(\textrm{det}A)^{-1}\textrm{det}(B-DA^{-1}C).
\end{equation*}

\subsection{Left and right derivatives}
The rigourous definition for these operations are given in  \citep{REF6}. Here we will give only
an example. Let us give a function $f(\xi_1,\xi_2)$ of two grassmannian odd variables $\xi_1,\,\xi_2$ of the form 
\[
f(\xi_1,\xi_2)=\xi_1\,\xi_2.
\]
Let us define the right or right derivative of $f$ with respect to $\xi_1$:
\begin{eqnarray*}
\frac{\overset{\rightarrow}{\partial }f}{\partial\xi_1}&=&\xi_2 \quad \textrm{\quad left derivative}\\
\frac{\overset{\leftarrow}{\partial }f}{\partial\xi_1}&=&-\xi_2 \quad \textrm{\quad right derivative}\\
\end{eqnarray*}
\invcomma{Somehow} roughly speaking in the right derivative it is as if we had put $\overset{\leftarrow}{\partial }f/\partial \xi_1$ to 
the right of the function so that $\overset{\leftarrow}{\partial }f/\partial \xi_1$ has to pass through $\xi_2$ in order to act on $\xi_1$.
In going through $\xi_2$ it acquires a minus sign because $\xi_1$ and $\xi_2$ anticommute. On grassmannian spaces  we can also define the concept of integration. All the details are given in  \citep{REF6}. The few things we need in this paper were already indicated in the body of the paper 
and will not be repeated in this appendix. 

 %%%%%%%%%%%%%%%%%%%%%%%%%%%%%%%%%%%%%%%%%%%%%%%%
%%%%%%%%%%%%%%%%%%%%%%%%%%%%%%%%%%%%%%%%%%%%%%%%
%%%%%%%%%%%%%%%%%%%%%%%%%%%%%%%%%%%%%%%%%%%%%%%%
\newpage
\section{\label{Appendix:B}}
In this appendix we will give details of the calculations of the vierbein in the CPI case.\\
Using the expression Eq.~(\ref{eq:41-1}) for the $E^M_{\,A}$ we get 
\[
D_t=\partial_M E^M_{\,\,t}=a\,\partial_t+\alpha\,\partial_{\theta}+\beta\,\partial_{\bar\theta},
\]
so 
\begin{equation*}
(D_tQ)(D_tQ)=a^2\,\partial_tQ\,\partial_tQ+2\,a\,\alpha\,\partial_tQ\,\partial_{\theta}Q
+2\,a\,\beta,\partial_tQ\,\partial_{\bar\theta}Q+\alpha\,\beta\,\partial_{\theta}Q\,\partial_{\bar\theta}Q.
\end{equation*}
Using the expression  above, it is easy to prove that a choice of parameters for which the second of Eq.~(\ref{eq:42-1}) holds is the following one:  
\begin{equation}
a=\pm1,\quad \alpha=\beta=0.
\label{eq:42-3}
\end{equation}
So the supervierbein in Eq.~(\ref{eq:41-1}) takes the form 
\begin{equation}
E^M_{\,A}=\begin{pmatrix}
\pm 1&0&0\\
\gamma&b&c\\
\delta&d&e\\
\end{pmatrix}
\label{eq:43-1}.
\end{equation} 
Next we have to impose the first of the condition Eq.~(\ref{eq:42-1}) using the definition of superdeterminant given in \citep{REF6} or  \ref{Appendix:A} of this paper and applied to Eq.~(\ref{eq:43-1}):
\begin{equation}
\textrm{sdet} E^M_{\,A}=1 \Longrightarrow \textrm{det}
\begin{pmatrix}
b&c\\
d&c
\end{pmatrix}
\Longrightarrow b\,e-c\,d=\pm 1.
\label{eq:43-2}
\end{equation}
The quantity $b,c,d,e$ are even so they have the form 
\begin{equation}
\begin{matrix}
b\equiv b_B+b_S\,\bar\theta\theta\\
c\equiv c_B+c_S\,\bar\theta\theta\\
d\equiv d_B+d_S\,\bar\theta\theta\\
e\equiv e_B+e_S\,\bar\theta\theta,
\end{matrix}
\label{eq:43-3}
\end{equation}
where $b_B,c_B,d_B,e_B$ are the  ``bodies" of the numbers while $b_S,c_S,d_S,e_S$
are called the ``souls" of the numbers. For details about these numbers see \ref{Appendix:A} of this paper and consult ref.\citep{REF6}. Using Eq.~(\ref{eq:43-3}) the relation (\ref{eq:43-2}) gives the two equations
\begin{equation}
\left\{
\begin{matrix}
b_B\,e_B-c_B\,d_B=\pm 1\\
b_S\,e_B+b_B\,e_S-c_S\,d_B-c_B\,d_S=0
\end{matrix}
\right..
\label{eq:44-1}
\end{equation}
A set of two solutions has the form 
\begin{eqnarray}
&(1)&\, e_B=0,\,c_B=\mp\frac{1}{d_B},\,c_S=\frac{\mp d_S+d_B\,b_B\,e_S}{d_B^2}\\
&(2)&\,b_B=\frac{\pm 1+c_B\,d_B}{e_B},\,b_S=\frac{\mp e_S-c_B\,d_B\,e_S}{e_B^2}+\frac{c_B\,d_S\,e_B+c_S\,d_B\,e_B}{e_B^2}\nonumber \label{eq:44-2}
\end{eqnarray}
It is a long but easy calculation to build the inverse \citep{REF6} of the matrix $E^M_{\,A}$, the result is
\begin{equation}
\begin{pmatrix}
\pm 1&0&0\\
\pm c\,\delta\mp e\,\gamma&\pm e&\mp c\\
\pm d\,\gamma\mp b\,\delta&\mp d&\pm b
\end{pmatrix}.
\end{equation}
So now we have all the matrix elements to build the kinetic term and the superdeterminat. The number of free elements that we know in Eq.~(\ref{eq:43-1}) is $12$ because each $b,c,d,e,\gamma,\delta$ is made of two numbers either the body and the soul of the even ones or for the odd elements, like $\gamma=\gamma_{\theta}\,\theta+\gamma_{\bar\theta}\,\bar\theta$, they are the coefficient of $\theta$ and of $\bar \theta$. We have  $2$ constraints in Eq.~(\ref{eq:44-1}), so the number of free variables is $10$ and we choose them to be real.  Considering the gauge freedom the careful reader may envision the following problem. We said that building the CPI or the QPI is ``like a gauge fixing''. For the CPI this ``gauge fixing '' is given by the constraints of Eq.~(\ref{eq:42-1}). One important thing to check is that the  `` gauge fixing'' Eq.~(\ref{eq:42-1}) does not fix more parameters than those allowed by the gauge freedom. This is not so and we prove it below. 
Our diffeomorphism, Eq.~(\ref{eq:34-2}), can be explicitly written as: 
\begin{eqnarray}
\label{eq:63-1bis}
\delta t&=&A(t)+\widetilde \alpha(t)\theta+\widetilde\beta(t)\bar\theta+\beta(t)\theta\bar\theta\nonumber\\
\delta \theta&=&\widetilde\gamma(t)+C(t)\theta+D(t)\bar\theta+\epsilon(t)\theta\bar\theta\\
\delta \bar\theta&=&\widetilde\delta(t)+F(t)\theta+G(t)\bar\theta+\xi(t)\theta\bar\theta,\nonumber
\end{eqnarray}
where the latin symbols are real even numbers and the greek ones are odd number functions only of $t$ and not of $\theta$ and $\bar\theta$.
So in (\ref{eq:63-1bis}) we have 12 parameters. The general vierbein has the form
\begin{equation}
E^M_A(z)=\begin{pmatrix}
a&\alpha&\beta\\
\gamma&b&c\\
\delta&d&e
\end{pmatrix}
\label{eq:63-2}
\end{equation}
and it contains 18 variables because each $a,\alpha,\dots, e$ are made of two entries. This vierbein transforms in the following manner under (\ref{eq:63-1bis}) or (\ref{eq:34-2}):
\begin{equation}
E'^M_A(z)=\frac{\overset{\rightarrow}{\partial }z'_B}{\partial z_A}E^M_B(z').
\label{eq:63-3}
\end{equation}
If we were able to fully exploit the 12-parameter gauge freedom of (\ref{eq:63-1bis}), we could reduce the 18-variables of $E^M_A$ to just six. In the CPI the vierbein that we use is: 
\begin{equation}
E^M_A(CPI)=\begin{pmatrix}
\pm 1& 0&0\\
\gamma&b&c\\
\delta&d&e 
\end{pmatrix}
\end{equation} 
so we have 12 parameters minus the 2 constraints (\ref{eq:44-1}) and this brings down to 10 parameters that are more than 6. So we have done only a partial gauge fixing.
\newpage
\section{\label{Appendix:C}}
In this appendix we will give details of the calculation of the vierbein for the QPI. \\For the QPI we will get a vierbein of the form  
\begin{equation}
E^M_A(QPI)=\begin{pmatrix}
 1+a_s \theta\bar\theta& \alpha&\beta\\
\gamma&b&c\\
\delta&d&e 
\end{pmatrix}, 
\end{equation} 
which has $17$ parameters minus $2$ constraints (that we will see later on) bringing the total free parameters down to $15$. Moreover we will choose $\alpha=\beta=0$ like in the CPI, so we will come down to $11$ parameters, which again is more than $6$. This is consistent with considering our procedure as a partial gauge fixing. This would not be so if the procedure would bring the number of free parameters to less than $6$ both in the CPI and in the QPI. \par
Let us now build the vierbein for the quantum case that is the QPI of Eq.~(\ref{eq:2}). In this case the determinant of the vierbein $E=\textrm{sdet}(E^A_{\,M})$ has to be  
\begin{equation}
E=-i\frac{\bar\theta\theta}{\hbar}
\label{eq:48-1}.
\end{equation}
This number has a body equal to zero and, as explained in \citep{REF6}, and in \ref{Appendix:A}, it does not admit an inverse $E^{-1}$. The inverse would be the determinant of the elements $E^M_{\,A}$ which eneter the kinetic piece of the action like it happened in the CPI (see Eq.~(\ref{eq:38-2})). So these elements cannot be built if we stick to the condition Eq.~(\ref{eq:48-1}). The way out is is to add a small ``regulating'' body $\epsilon$ to Eq.~(\ref{eq:48-1}) so that the determinant can be inverted. This  \invcomma{regularized} determinant is 
\begin{equation}
E^{\textrm{reg}}=\epsilon -i\frac{\bar\theta\theta}{\hbar}
\label{eq:49-1}.
\end{equation}
The inverse can now be built \citep{REF6} and it is 
\begin{equation}
E^{-1}=\frac{1}{\epsilon}+\frac{1}{\epsilon^2}\,\frac{\bar\theta\theta}{\hbar}
\label{eq:49-2}.
\end{equation}
We will now go on to find for the QPI the analog of the two constraints of Eq.~(\ref{eq:42-1}). Let us insert the regularized $E^{\textrm{reg}}$ of Eq.~(\ref{eq:49-1}) into the action of the QPI written in Eq.~(\ref{eq:29-2}) once we have integrated out the $P$. Moreover let us keep only the kinetic piece:
\begin{equation}
S^{\textrm{reg}}_{QPI}=i\int dt d\theta d\bar\theta\left(\epsilon -i\frac{\bar\theta\theta}{\hbar}\right)\left(\frac{1}{2}D_tQ\,D_tQ\right)
\label{eq:49-3}.
\end{equation}
Performing the products above,we get:
\begin{equation}
S^{\textrm{reg}}_{QPI}=i\,\epsilon\int dt d\theta d\bar\theta\left(\frac{1}{2}D_tQ\,D_tQ\right)+\frac{i}{\hbar}\int dtd\theta d\bar\theta\,\bar\theta\theta \left(\frac{1}{2}D_tQ\,D_tQ\right)
\label{eq:50-1}.
\end{equation}
The first term goes to zero in the true quantum-case because in this case $\epsilon\to 0$. So we will work out only the second term in Eq.~(\ref{eq:50-1}) using the general form of the vierbein written in Eq.~(\ref{eq:41-1}) and using the expression Eq.~(\ref{eq:38-2}) for the covariant derivative. The second term in Eq.~(\ref{eq:50-1}) turns out to be 
 \begin{equation}
\frac{1}{\hbar}\int dt\,d\theta\, d\bar\theta \,\frac{1}{2}\left(a\,\partial_t Q+d\,\partial_t Q+\beta\,\partial_{\bar\theta} Q\right)^2\,\bar\theta\theta.
\label{eq:50-2}.
\end{equation}
As the $\alpha$ and $\beta$ are odd and get bultiplied by $\bar\theta\theta$ the only term which survives is 
 \begin{equation}
\frac{1}{\hbar}\int dt\,d\theta\, d\bar\theta \,\frac{1}{2}\,a^2\,(\partial_t Q)^2\,\bar\theta\theta
\label{eq:50-3}.
\end{equation}
Differently than in the classical case of Eq.~(\ref{eq:42-3}), note that in the quantum case
$a$ is an even element made of a body $a_B$ and a soul $a_S$. 
So Eq.~(\ref{eq:50-3}) turns out to be 
 \begin{equation}
\frac{1}{\hbar}\int dt\,d\theta\,d\bar\theta\cfrac{1}{2}\,\left(a_B^2+2\,a_B\,a_S\,\bar\theta\theta\right)\bar\theta\theta\partial_tQ\partial_tQ=\frac{1}{2\,\hbar}\int \,a_B^2\,(\partial_t q)(\partial_t q)
\label{eq:51-1}.
 \end{equation}
%%%%%%%%%%%%%%%%%%%%%%%%%%%%%%%%%%%%%%%%%%%%%%%%
%%%%%%%%%%%%%%%%%%%%%%%%%%%%%%%%%%%%%%%%%%%%%%%%
%%%%%%%%%%%%%%%%%%%%%%%%%%%%%%%%%%%%%%%%%%%%%%%%
where $q$ is the first component of $Q$ like in Eq.~(\ref{eq:24-1}) and we have omitted the indices for simplicity. 
In order to get the usual kinetic piece of quantum mechanics in Eq.~(\ref{eq:51-1})  we need to have:
\begin{equation}
a_B=\pm 1,
\label{eq:51-2}
\end{equation}
while $a_S$ is free. Next we have to impose the conditions Eq.~(\ref{eq:49-2}) and Eq.~(\ref{eq:51-2}) on the determinant, i.e.
\begin{equation}
\textrm{sdet}\begin{pmatrix}
\pm 1+a_S \,\bar\theta\theta&\alpha&\beta\\
\gamma&b&c\\
\delta&d&e
\end{pmatrix}=\frac{1}{\epsilon}+\frac{i}{\epsilon}\frac{\bar\theta\theta}{\hbar}.
\label{eq:51-3}
\end{equation}
Working out the sdet on the L.H.S. of  Eq.~(\ref{eq:51-3}) using the usual rules given in \citep{REF6}, we get 
\begin{eqnarray}
\label{eq:52-1}
&\textrm{sdet}&
\begin{pmatrix}
\pm 1+a_S \,\bar\theta\theta&\alpha&\beta\\
\gamma&b&c\\
\delta&d&e
\end{pmatrix}\\&=&\left[
\pm 1+a_S \,\bar\theta\theta-\begin{pmatrix}
\alpha&\beta
\end{pmatrix} \begin{pmatrix}
b&c\\
d&e
\end{pmatrix}^{-1}\begin{pmatrix}
\gamma\\ \delta
\end{pmatrix}
\right]\cdot \textrm{det}^{-1}\begin{pmatrix}
b&c\\
d&e
\end{pmatrix}.
\nonumber
\end{eqnarray}
Let us now simplify things by introducing some new symbols $p,q,r$ defined as:
\begin{eqnarray}
p\,\bar\theta\theta&\equiv& \begin{pmatrix}
\alpha&\beta 
\end{pmatrix}\begin{pmatrix}
b&c\\
d&e
\end{pmatrix}^{-1}\begin{pmatrix}
\gamma\\ \delta
\end{pmatrix}\nonumber\\
q+r\,\bar\theta\theta&\equiv&  \textrm{det}^{-1}\begin{pmatrix}
b&c\\
d&e
\end{pmatrix}.
\label{eq:52-2}
\end{eqnarray}
The powers of $\theta,\bar\theta$ present on the L.H.S.of Eq.~(\ref{eq:52-2}) can be easily understood by remembering the powers of  $\theta,\bar\theta$ present in the even and odd elements. 
Using Eq.~(\ref{eq:52-2}) the relation Eq.~(\ref{eq:52-1}) can be written as 
\begin{equation}
\textrm{sdet}\begin{pmatrix}
\pm 1+a_S \,\bar\theta\theta&\alpha&\beta\\
\gamma&b&c\\
\delta&d&e
\end{pmatrix}=\pm q+(a_S\,q-p\,q\pm r)\bar\theta\theta
\label{eq:52-3}.
\end{equation} 
Combining  Eq.~(\ref{eq:52-3}) with Eq.~(\ref{eq:51-3}) we get 
\begin{equation}
\left\{\begin{matrix}
&&q=\pm \cfrac{1}{\epsilon}\\
&&a_sq-p\,q\pm r= \cfrac{i}{\epsilon^2\,\hbar}
\end{matrix}\right.
\label{eq:53-1}.
\end{equation}
 which  can be combined to give 
 \begin{equation}
 \pm\left(\frac{a_S}{\epsilon}-\frac{p}{\epsilon}+r\right)=\frac{1}{\epsilon^2\,\hbar}
\label{eq:53-2}.
 \end{equation}
From the second equation in (\ref{eq:52-2}) we get that the matrix 
\begin{equation}
D\equiv\begin{pmatrix}
b&c\\d&e
\label{eq:53-3}
\end{pmatrix}
\end{equation}
must be invertible and the determinant of the inverse must be equal to the L.H.S. of the following equation: 
\begin{equation}
q+r\,\bar\theta\theta=\pm\frac{1}{\epsilon}+r\,\bar\theta\theta.
\label{eq:53-4}
\end{equation}
The R.H.S. of Eq.~(\ref{eq:53-4}) is otained from the first of Eq.~(\ref{eq:53-1}). From the determinant of the inverse we can get 
 the determinant of $D$ which from Eq.~(\ref{eq:53-4}) turns out to be 
 \begin{equation}
 \textrm{det}D=\pm \epsilon-\epsilon^2\,r\,\bar\theta\theta
\label{eq:53-5}.
\end{equation} 
The determinant of $D$ is equal to $(b\,e-c\,d)$ so Eq.~(\ref{eq:53-5}) becomes 
 \begin{equation}
b\,e-c\,d=\pm \epsilon-\epsilon^2\,r\,\bar\theta\theta,
\label{eq:54-1}
\end{equation} 
which is equal to 
\begin{equation}
\label{eq:54-2}
(b_B+b_S\,\bar\theta\theta)(e_B+e_S\,\bar\theta\theta)-(c_B+c_S\,\bar\theta\theta)(d_B+d_S\,\bar\theta\theta)=\pm \epsilon-\epsilon^2\,r\,\bar\theta\theta
\end{equation}
and comparing equal powers of $\theta$ and $\bar\theta$ we get that Eq.~(\ref{eq:54-2})
is equivalent to the following two equations
\begin{eqnarray}
\left\{
\begin{matrix}
b_B\,e_B-c_B\,d_B=\pm \epsilon\\
b_S\,e_B+b_B\,e_S-c_S\,d_B-c_B\,d_S=-\epsilon^2\,r.
\end{matrix}
\right.
\label{eq:54-3}
\end{eqnarray}
The first equation is a true constraint equation while the second one relates the parameter $r$ to the variables $b,c,d,e$.
From Eq.~(\ref{eq:52-2}) we can also obtain the  detail expression  of $p$ in terms  of the entries of the vierbein. A long calculation leads to the following equation 
\begin{eqnarray}
p=\pm\frac{1}{\epsilon}\left(
\alpha_{\bar\theta}\,\gamma_{\theta}\,e_B-\gamma_{\bar\theta}\,\alpha_{\theta}\,e_B-\alpha_{\bar\theta}\,\delta_{\theta}\,c_B+\alpha_{\theta}\,\delta_{\bar\theta}\,c_B\right.\nonumber\\
\left.+\beta_{\theta}\,\gamma_{\bar\theta}\,d_B-\beta_{\bar\theta}\,\gamma_{\theta}\,d_B+\beta_{\bar\theta}\,\delta_{\theta}\,b_B-\beta_{\theta}\,\delta_{\bar\theta}\,b_B\right).
\label{eq:55-1}
\end{eqnarray}
Inserting Eq.~(\ref{eq:55-1}) and the second of Eq.~(\ref{eq:54-3}) into the second of Eq.~(\ref{eq:53-1}) we get a constraint among the $a,b,c,d,\alpha,\beta,\gamma,\delta$. This constraint together with the first of Eq.~(\ref{eq:54-3}) provides the two QPI constraints analog to the two of the CPI of Eq.~(\ref{eq:44-1}) but much more complicated.
In order to simplify things let us choose $\alpha=\beta=0$ like in the CPI case. This choice, once  inserted in Eq.~(\ref{eq:55-1}), gives $p=0$. Using this inside Eq.~(\ref{eq:53-2}) we get 
 \begin{equation}
 \frac{a_S}{\epsilon}+r=\pm \frac{i}{\epsilon^2\,\hbar}
\label{eq:56-1}
\end{equation}  
and using for $r$ the expression in the second equation of Eq.~(\ref{eq:54-3}) we get from Eq.~(\ref{eq:56-1}) the following constraint 
\begin{equation}
b_S\,e_B+b_B\,e_S-c_S\,d_B-c_B\,d_S=\mp \frac{i}{\hbar}+\epsilon\,a_S.
\label{eq:56-2}
\end{equation}
This together with the first relation in Eq.~(\ref{eq:54-3}) are the two constraints for the QPI analog to the two for the CPI in Eq.~(\ref{eq:44-1}). Let us write together those of the QPI
\begin{equation}
\label{eq:56-3}
\begin{lcases}
&b_B\,e_B-c_B\,d_B=\pm \epsilon\hspace{3cm}\\
&b_S\,e_B+b_B\,e_S-c_S\,d_B-c_B\,d_S=\mp \cfrac{i}{\hbar}+\epsilon\,a_S.
\end{lcases}
\end{equation}
We can find some solutions of these equations like for example
\begin{eqnarray}
&(1)&\quad e_B=0,\quad c_B=\mp\frac{1}{d_B},\quad c_S=\frac{\pm \epsilon\,d_S+d_B\,b_B\,a_S\pm \frac{i}{\hbar}a_S\,d_B-\epsilon\,a_S\,d_B}{d_B^2}\nonumber \\
&(2)&\,b_B=\frac{\pm \epsilon+c_B\,d_B}{e_B},\quad b_S=\frac{\mp \epsilon\,e_S-c_B\,d_B\,b_S+c_B\,d_S\,e_B}{e_B^2}\nonumber\\ 
&&+\frac{c_B\,d_B\,e_B\mp \frac{i}{\hbar}e_B+\epsilon\,a_S\,a_B}{e_B^2}.
\label{eq:57-1}
\end{eqnarray}
\newpage
%%%%%%%%%%%%%%%%%%%%%%%%%%%%%%%%%%%%%%%%%%%%%%%%%%%%%%%%%%%
%%%%%%%%%%%%%%%%%%%%%%%%%%%%%%%%%%%%%%%%%%%%%%%%%%%%%%%%%%%
\section{\label{Appendix:D}}
Here we will give details for the construction of the metric in the QPI case.
Let us start with the most general vierbein:
\begin{equation}
E_A^{\,\,M}=\begin{pmatrix}
a&\alpha&\beta\\
\gamma&b&c\\
\delta&d&e
\end{pmatrix}
\label{eq:62-1}
\end{equation}  
and later on we will insert the quantum constraints Eq.~(\ref{eq:54-3}) in the associated matrix. 
Using the relation Eq.~(\ref{eq:58-2}) between vierbein and metric we get that the metric associated to the general vierbein Eq.~(\ref{eq:62-1}) has the form:
\begin{equation}
\begin{pmatrix}
1-2\gamma\,\delta+2\,\bar\theta\theta\,a_Ba_S&d\,\gamma-b\delta+\alpha a_B&e\,\gamma-c\,\delta +\beta\, a_B\\
d\,\gamma-b\,\delta+\alpha \,a_B&0&b e-c d+\alpha\,\beta\\
e\,\gamma -c\,\delta+\beta\, a_B&c d-b e -\alpha \,\beta&0
\end{pmatrix}.
\label{eq:63-1}
\end{equation}
%%%%%%%%%%%%%%%%%%%%%%%%%%%%%%%%%%%%%%%%%%%%%%%%
%%%%%%%%%%%%%%%%%%%%%%%%%%%%%%%%%%%%%%%%%%%%%%%%
%%%%%%%%%%%%%%%%%%%%%%%%%%%%%%%%%%%%%%%%%%%%%%%%
 Let us now rewrite the metric using the $\pi_i$ introduced in  Eq.~(\ref{eq:60-1}) and Eq.~(\ref{eq:60-2}). From the definition of $\pi_5$ in Eq.~(\ref{eq:60-2}) it is easy to prove that 
 \[
\gamma\,\delta=\pi_5\,\bar\theta\theta 
 \]
 so the element $g^{11}$ of Eq.~(\ref{eq:63-1}) can be written as 
\begin{eqnarray}
1-2\gamma\,\delta+2\,\bar\theta\theta\,a_Ba_S&=&1-2(\pi_5-a_Ba_S)\bar\theta\theta\nonumber\\&\equiv&1-\pi_5^Q\,\bar\theta\theta
\label{eq:64-1}
\end{eqnarray} 
 where we have defined a new quantity $\pi^Q_5$ as 
 \begin{equation}
\pi^Q_5\equiv \pi_5-a_Ba_S.
 \label{eq:64-2}
 \end{equation}
The index \invcomma{Q} is to indicate that these are objects related to the QPI. The element  $g^{12}$ of Eq.~(\ref{eq:63-1}) can be written as 
 \begin{eqnarray}
 d\,\gamma-b\,\delta+\alpha \,a_B&=&-\pi_3\,\theta-\pi_4\,\bar\theta+\alpha_{\theta}\,a_B\,\theta+\alpha_{\bar \theta}\,a_B\,\bar\theta\nonumber\\
 &=&-(\pi_3-\alpha_{\theta}a_B)\theta-(\pi_4-\alpha_{\bar\theta}a_B)\bar\theta\nonumber\\
 &\equiv&-\pi_3^Q\,\theta-\pi^Q_4\,\bar\theta
  \label{eq:64-3}
 \end{eqnarray}
where  
 \begin{eqnarray}
 \pi_3^Q&\equiv&\pi_3-\alpha_{\theta}a_B\nonumber\\
\pi_4^Q&\equiv&\pi_4-\alpha_{\bar\theta}a_B. 
  \label{eq:64-4}
 \end{eqnarray}
 The element $g^{13}$ of Eq.~(\ref{eq:63-1}) can be written as 
\begin{eqnarray}
e\,\gamma -c\,\delta+\beta\, a_B&=&\pi_1\,\theta+\pi_2\,\bar\theta+\beta \,a_B\nonumber\\ 
&=&\pi_1\,\theta+\pi_2\,\bar\theta+\beta_{\theta}\,\theta\,a_B+\beta_{\bar\theta}\,\bar\theta\,a_B\nonumber\\
&=&(\pi_1+\beta_{\theta}a_B)\theta+(\pi_2+\alpha_{\bar\theta}a_B)\bar\theta\nonumber\\
&\equiv& \pi^Q_1\,\theta+\pi^Q_2\,\bar\theta
  \label{eq:65-1}
\end{eqnarray} 
where  
 \begin{eqnarray}
 \pi_1^Q&\equiv&\pi_1+\beta_{\theta}a_B\nonumber\\
\pi_2^Q&\equiv&\pi_2+\beta_{\bar\theta}a_B. 
  \label{eq:65-2}
 \end{eqnarray}
 Next let now examine the term $g^{23}$ of Eq.~(\ref{eq:63-1}) 
\begin{eqnarray}
  \label{eq:65-3}
b e-c d+\alpha\,\beta&=&(b_B\,e_B-c_B\,d_B)+\\
&+&(b_S\,e_B+b_B\,e_S-c_S\,d_B-c_B\,d_S+\alpha_{\bar\theta}\,\beta_{\theta}-\beta_{\bar\theta}\,\alpha_{\theta})\,\bar\theta\theta.\nonumber
\end{eqnarray} 
When $a_B\neq\pm 1$ the first of relation  (\ref{eq:54-3}) and  (\ref{eq:53-2}) will turn into the following two relations:
\begin{equation}
\left\{\begin{matrix}
 b_B\,e_B-e_B\,d_B&=&a_B\,\epsilon\\
 &&\\
\cfrac{a_Ba_S}{\epsilon}-\cfrac{a_Bp}{\epsilon}+a_Br&=&\cfrac{i}{\epsilon^2\,\hbar}
\end{matrix}\right..
  \label{eq:66-1}
\end{equation}
The first of relations (\ref{eq:53-1}) will turn into 
\[q=\frac{a_B}{\epsilon}\]
while the relation Eq.~(\ref{eq:55-1}) for $p$ becomes
\begin{eqnarray}
p&=&\frac{a_B}{\epsilon}\left[\alpha_{\bar\theta}(\gamma_{\theta}\,e_B-\delta_{\theta}\,c_B)+\alpha_{\theta}(\delta_{\bar\theta}\,c_B-\gamma_{\bar\theta}\,e_B)+\right. \nonumber\\
 && \left.  
+\beta_{\theta}(\gamma_{\bar\theta}\,d_B-\delta_{\bar\theta}\,b_B)+\beta_{\bar{\theta}}(\delta_{\theta}\,b_B-\gamma_{\theta}\,d_B)\right]
 \nonumber\\
&=& \frac{a_B}{\epsilon}\left(\alpha_{\bar \theta}\,\pi_1-\alpha_{\theta}\,\pi_2-\beta_{\theta}\,\pi_4+\beta_{\bar\theta}\,\pi_3\right)
%\left. 
  \label{eq:66-2}
\end{eqnarray}
Multiplying  the second equation of (\ref{eq:66-1}) by $\epsilon^2/a_B$ we get 
\[
\epsilon\,a_S-\epsilon\,p+\epsilon^2\,r=\cfrac{i}{\hbar\,a_B}
\]
which, using Eq.~(\ref{eq:66-2}), implies
\begin{eqnarray}
-\epsilon^2\,r&=&\epsilon\,a_S-\epsilon\,p-\frac{i}{\hbar\,a_B}=\\
&=&\epsilon\,a_S-\frac{i}{\hbar\,a_B}-a_B(\alpha_{\bar \theta}\,\pi_1-\alpha_{\theta}\,\pi_2-\beta_{\theta}\,\pi_4+\beta_{\bar\theta}\,\pi_3)\nonumber.
  \label{eq:66-3}
\end{eqnarray}
Let us now remember the second relation of Eq.~(\ref{eq:54-3})
\begin{equation}
b_S\,e_B+b_B\,e_S-c_S\,d_B-c_B\,d_S=-\epsilon^2\,r.
  \label{eq:67-1}
\end{equation}
Note that the L.H.S. of this equation are exactly the first four terms of the soul of $b\,e-c\,d+\alpha\,\beta$ in Eq.~(\ref{eq:65-3}). Replacing them with the expression of $-\epsilon^2\,r$, which appear on the L.H.S. of Eq.~(\ref{eq:66-3}), we get that the soul of $b\,e-c\,d+\alpha\,\beta$ is equal to 
\begin{eqnarray}
\epsilon\,a_S&-&\frac{i}{\hbar\,a_B}+a_B(\alpha_{{\theta}}\,\pi_2-\alpha_{\bar \theta}\,\pi_1+\beta_{\theta}\,\pi_4-\beta_{\bar\theta}\,\pi_3)+\nonumber\\&+&\alpha_{\bar\theta}\,\beta_{\theta}-\alpha_{\theta}\,\beta_{\bar\theta}\equiv \pi_6^Q
  \label{eq:67-2}.
\end{eqnarray}
In the equation above the soul of $b\,e-c\,d+\alpha\,\beta$ has been set equal to $\pi_6^Q$.

Going now back to  Eq.~(\ref{eq:65-3}) and using for its body the first constraint of Eq.~(\ref{eq:66-1}) we get 
\begin{equation}
b\,e-c\,d+\alpha\,\beta=a_B\,\epsilon+\bar\theta\theta\,\pi_6^Q
  \label{eq:67-3}.
\end{equation}
We have now all elements to write down the metric with all constraints implemented 
\begin{equation}
g^{MN}=
\begin{pmatrix}
1-2\,\pi_5^Q\,\bar\theta\theta&-\pi_3^Q\,\theta-\pi_4^Q\,\bar\theta&\pi_1^Q\,\theta+\pi_2^Q\,\bar\theta\\
-\pi_3^Q\,\theta-\pi_4^Q\,\bar\theta&0&a_B\,\epsilon+\pi^Q_6\bar\theta\theta\\
\pi_1^Q\theta+\pi_2^Q\bar\theta&-a_B\,\epsilon-\pi_6^Q\,\bar\theta\theta&0
\end{pmatrix}.
  \label{eq:68-1}.
\end{equation}
If we define a new variable $\pi_7^Q$ as 
\begin{equation}
\pi_7^Q=a_B\,\epsilon
  \label{eq:68-2},
\end{equation}
the metric Eq.~(\ref{eq:68-1}) can be written using only  $\pi_i^Q$ variables as follows:
\begin{equation}
g^{MN}=
\begin{pmatrix}
1-2\,\pi_5^Q\,\bar\theta\theta&-\pi_3^Q\,\theta-\pi_4^Q\,\bar\theta&\pi_1^Q\,\theta+\pi_2^Q\,\bar\theta\\
-\pi_3^Q\,\theta-\pi_4^Q\,\bar\theta&0&\pi_7^Q+\pi^Q_6\bar\theta\theta\\
\pi_1^Q\theta+\pi_2^Q\bar\theta&-\pi_7^Q-\pi_6^Q\,\bar\theta\theta&0
\end{pmatrix}.
  \label{eq:68-3}
\end{equation}
Let us summarize the various quantities we have introduced:  
\begin{equation}
\begin{lcases}
\pi_1^Q&\equiv\pi_1+\beta_{\theta}\,a_B\\
\pi_2^Q&\equiv\pi_2+\beta_{\theta}\,a_B\\
\pi_3^Q&\equiv\pi_3-\alpha_{\theta}\,a_B\\
\pi_4^Q&\equiv\pi_4-\alpha_{\bar\theta}\,a_B\\
\pi_5^Q&\equiv\pi_5-a_B\,a_S\\
\pi_6^Q&\equiv\epsilon\,a_S-\cfrac{i}{\hbar\,a_B}+a_B(\alpha_{\theta}\,\pi_2-\alpha_{\bar\theta}\pi_1+\beta_{\theta}\pi_4-\beta_{\bar\theta}\pi_3)+\alpha_{\bar\theta}\beta_{\theta}-\alpha_{\theta}\beta_{\bar\theta}\\
\pi_7^Q&\equiv a_B\epsilon
\end{lcases}
 \label{eq:69-1}
\end{equation}
 %%%%%%%%%%%%%%%%%%%%%%%%%%%%%%%%%%%%%%%%%%%%%%%%
%%%%%%%%%%%%%%%%%%%%%%%%%%%%%%%%%%%%%%%%%%%%%%%%
%%%%%%%%%%%%%%%%%%%%%%%%%%%%%%%%%%%%%%%%%%%%%%%%
Let us now count the number of free parameters. $\alpha$ and $\beta$ do not enter any of the constraints in  Eq.~(\ref{eq:54-3}) so they are free. In order to simplify things we can put them by hand equal to zero like in the CPI and we suggested this already after  Eq.~(\ref{eq:55-1}). Moreover we should remember that $a_B=\pm 1$ as proved in  Eq.~(\ref{eq:51-2}) but differently than the classical case $a_S$ is a free parameter. So with this choice  Eq.~(\ref{eq:69-1}), becomes 
\begin{equation}
\begin{lcases}
\pi_1^Q&=\pi_1\\
\pi_2^Q&=\pi_2\\
\pi_3^Q&=\pi_3\\
\pi_4^Q&=\pi_4\\
\pi_5^Q&=\pi_5\mp a_S\\
\pi_6^Q&=\epsilon\,a_S\mp\cfrac{i}{\hbar}\\
\pi_7^Q&=\pm\epsilon
\end{lcases}
 \label{eq:70-1}
\end{equation}
So the \invcomma{quantum} metric depend on $5$ parameters $\pi_1^Q,\pi_2^Q,\pi_3^Q,\pi_4^Q$ and $a_S$ while the classical one only on $4$. The reader may object that also $\alpha$ and $\beta$ were free and should be counted. He is right. Anyhow as we put them equal to zero both in the CPI and the QPI and so the difference in the numbers of free parameters remains one between QM and CM Let us look at the vierbein. For the CPI we have $10$ free parameters, while in the QPI will be $11$ because we have $a_S$ as extra variable. If we had not put $\alpha=\beta=0$ we would have $14$ parameters for the vierbein of the CPI and $15$ for the QPI. For the metrics instead, as the $\alpha$ and $\beta$ get incorporated into the $\pi^Q_i$ (see Eq.~(\ref{eq:69-1}))  the number of free parameters is $6$ for the QPI, while for the CPI we do not know because we should re-derive the metric keeping the $\alpha$ and $\beta$ different from zero.To finish this section let us explore the issue of wether we can recover the classical case from the \invcomma{regulated quantum} one without setting $\hbar\to 0$ but manipolating the $\epsilon$ parameter and the others.    
For sure we have to require that $\alpha=\beta=a_S=0$ which are the values set previously in the CPI. Moreover in the CPI we had the constraint
\begin{equation}
b\,e-c\,d=\pm 1
 \label{eq:72-1}
\end{equation}
while in the QPI we had (with $\alpha=\beta=0$) Eq.~(\ref{eq:67-3}):
\begin{equation}
b\,e-c\,d=a_B\,\epsilon+\bar\theta\theta\,\pi_6^Q.
 \label{eq:72-2}
\end{equation}
For  Eq.~(\ref{eq:72-2}) to be equal to  Eq.~(\ref{eq:72-1}), as we know that $a_B=\pm 1$, we have to require that $\epsilon\to 1$ and $\pi^Q_6=0$. Actually, remembering  the form of $\pi^Q_6$ present in Eq.~(\ref{eq:69-1}), we see that the following other form of $\pi_6^Q$ 
\begin{eqnarray}
\pi_6^Q&=&\epsilon\,a_S-\cfrac{i}{\hbar}\,a_B\,(1-\epsilon)+a_B(\alpha_{\theta}\,\pi_2-\alpha_{\bar\theta}\,\pi_1+\beta_{\theta}\,\pi_4+\nonumber\\&-&\beta_{\bar\theta}\,\pi_3)+\alpha_{\bar\theta}\,\beta_{\theta}-\alpha_{\theta}\,\beta_{\bar\theta}
 \label{eq:72-3}
\end{eqnarray}
has the same quantum limit  ($\epsilon\to 0$) as the one conained in Eq.~(\ref{eq:69-1}). So we can use Eq.~(\ref{eq:72-3}) in order to reproduce QM This new $\pi_6^Q$ has the feature that it goes to zero for $\epsilon=1$ (of course this has to be combined with the other things we require for CM: $a_S=\alpha=\beta=0$).  So in the limit $\epsilon \to 0$ we would get QM and in the limit $\epsilon\to 1$ we would get CM This is equivalent of having required that the determinant of the vierbein had the form 
\begin{equation}
E=\epsilon-i\,(1-\epsilon)\,\frac{\bar\theta\theta}{\hbar}.
 \label{eq:73-1}
\end{equation}
For $\epsilon\to 1$ we would get from  Eq.~(\ref{eq:73-1})
\[
E=\mathbb I
\]
which is the $C.P.I$ and for $\epsilon\to 0$ we would get 
\[
E=-i\,\frac{\bar\theta\theta}{\hbar},
\]
which is QM
For $\epsilon$ in between 0 and 1 we would get a family of models which are betwwen CM and QM and could interpolate all the mesoscopic physics. Before concluding this section let us provide the inverse of $g^{MN}$ of Eq.~(\ref{eq:68-3}). This quantity will be useful for the calculations provided in the next section. 
\begin{equation}
g_{MN}=\begin{pmatrix}
1-\frac{2\,\bar\theta\theta(\varphi^Q-\pi_5^Q\pi_7^Q)}{\pi_7^Q}&-\frac{\theta\,\pi_1^Q+\bar\theta\,\pi_2^Q}{\pi_7^Q}&-\frac{\theta\,\pi_3^Q+\bar\theta\,\pi_4^Q}{\pi_7^Q}\\
\frac{\theta\,\pi_1^Q+\bar\theta\,\pi_2^Q}{\pi_7^Q}&0&-\frac{\pi_7^Q+(\varphi^Q-\pi_6^Q)}{\pi_7^{Q^2}}\\
\frac{\theta\,\pi_3^Q+\bar\theta\,\pi_4^Q}{\pi_7^Q}&\frac{\pi_7^Q+\bar\theta\theta(\varphi^Q-\pi_6^Q)}{\pi_7^{Q^2}}&0
\end{pmatrix}.
 \label{eq:75-1}
\end{equation}
In the expression above to be compact we have defined 
\[
\varphi^Q\equiv\pi_2^Q\,\pi_3^Q-\pi_1^Q\,\pi_4^Q.
\]
Note that in the true quantum limit $\epsilon\to 0$ this metric is singular because $\pi_7^Q\to 0$.
%%%%%%%%%%%%%%%%%%%%%%%%%%%%%%%%%%%%%%%%%%%%%%%%%%%%%%%%%%%%%%%%
%%%%%%%%%%%%%%%%%%%%%%%%%%%%%%%%%%%%%%%%%%%%%%%%%%%%%%%%%%%%%%%%
%%%%%%%%%%%%%%%%%%%%%%%%%%%%%%%%%%%%%%%%%%%%%%%%%%%%%%%%%%%%%%%%
\newpage
\section{\label{Appendix:E}}
We will now calculate the Christofel symbols for the CPI leaving the body \invcomma{$a$} of the vierbein undetermined. The results, obtained assuming 
the $\pi_i$ independent on $t$ and using {\it Mathematica}, are the following ones: 
\begin{eqnarray}
\label{eq:B-1-1}
\Gamma_{t\theta}^t&=&\cfrac{\theta\pi_1(\pi_2-\pi_3)+\bar\theta(\pi_2(\pi_2+\pi_3)-2\pi_1\pi_4)}{2\,a}=-\Gamma_{\theta t}^t\nonumber\\
\Gamma_{t\theta}^\theta&=&\cfrac{\pi_2+\pi_3}{2}=\Gamma_{\theta t}^\theta=-\Gamma_{t \bar\theta}^{\bar\theta}=-\Gamma_{\bar\theta t}^{\bar\theta}\nonumber\\
\Gamma_{t\theta}^{\bar\theta}&=&-\pi_1=\Gamma_{\theta t}^{\bar\theta}\nonumber\\
\Gamma_{t\bar\theta}^t&=&\cfrac{\theta(-\pi_3(\pi_2+\pi_3)+2\pi_1\pi_4)+\bar\theta\,\pi_4(\pi_2-\pi_3)}{2\,a}=-\Gamma_{\bar\theta t}^t\\
\Gamma_{t\bar\theta}^\theta&=&\pi_4=\Gamma_{\bar\theta t}^\theta\nonumber\\
\Gamma_{\theta\bar\theta}^t&=&\cfrac{\pi_2-\pi_3}{2\,a}-2\,\bar\theta\theta(\pi_2-\pi_3)(\pi_2\pi_3-\pi_1\pi_4)=-\Gamma_{\bar\theta\theta}^t\nonumber\\
\Gamma_{\theta\bar\theta}^\theta&=&\cfrac{\theta(\pi_3(3\pi_2-\pi_3)-2\pi_1\pi_4)+\bar\theta\pi_4(\pi_2-\pi_3)}{2\,a}=-\Gamma_{\bar\theta\theta}^\theta\nonumber\\
\Gamma_{\theta\bar\theta}^{\bar\theta}&=&\cfrac{\theta(\pi_1(\pi_3-\pi_2)+\bar\theta(\pi_2(3\pi_3-\pi_2)-2\pi_1\pi_4))}{2\,a}=-\Gamma_{\bar\theta\theta}^{\bar\theta}.\nonumber
\end{eqnarray}
All the other Christofel symbols are equal to zero.
Similarly we can calculate the Christofel symbols for the QPI using the metric (\ref{eq:68-3}) and (\ref{eq:75-1}). 
In order to simplify the expression for the Christofel symbols and curvatures, we need to introduce the following quantity:  
\begin{equation}
\sigma_1\equiv\pi_2^Q\pi_3^Q-\pi_1^Q\pi_4^Q-\pi_5^Q\pi_7^Q.
\label{eq:B-1-1bis}
\end{equation}
Note that in the classical limit $\pi_i^Q\to \pi_i,\,i=1,\dots,4$ since  $\alpha_{\theta},\alpha_{\bar\theta},\beta_{\theta},\beta_{\bar\theta}\to 0$ and $\pi_7^Q\to a$ being $\epsilon\to 1$. Therefore $\sigma_1\to (\pi_2\pi_3-\pi_1\pi_4-\pi_5\,a_B)$, that in the classical limit is equal to zero. The result, via Mathematica \citep{REF14}, for parameters $\pi_i$ independent on $t$, turns out to be: 
\begin{eqnarray}
\Gamma_{tt}^t&=&\bar\theta\theta\cfrac{(\pi_3^Q-\pi_2^Q)}{\pi_7^Q}\,\sigma_1\nonumber\\
\Gamma_{tt}^\theta&=&-\bar\theta\,\sigma_1\nonumber\\
\Gamma_{t\theta}^t&=&\Gamma_{t\theta}^{t\; CPI}(\pi_i^Q)+\cfrac{\sigma_1}{\pi_7^Q}=-\Gamma_{\theta t}^t\nonumber\\
\Gamma_{t\theta}^\theta&=&\Gamma_{t\theta}^{\theta\; CPI}(\pi_i^Q)+\bar\theta\theta\,\cfrac{\pi_6^Q(\pi_2^Q+\pi_3^Q)+2\pi_3^Q\,\sigma_1}{2\,\pi_7^Q}=\Gamma_{\theta t}^\theta\nonumber\\
\Gamma_{t\theta}^{\bar\theta}&=&\Gamma_{t\theta}^{\bar\theta\; CPI}(\pi_i^Q)-\bar\theta\theta\,\cfrac{\pi_1^Q\,(\sigma_1+\pi_6^Q)}{\pi_7^Q}=-\Gamma_{\theta t}^{\bar\theta}\nonumber\\
\Gamma_{t\bar\theta}^{t}&=&\Gamma_{t\bar\theta}^{t\; CPI}(\pi_i^Q)-\cfrac{\sigma_1}{\pi_7^Q}=-\Gamma_{\bar\theta t}^t\nonumber\\
\Gamma_{t\bar\theta}^{\theta}&=&\Gamma_{t\bar\theta}^{\theta\; CPI}(\pi_i^Q)+\bar\theta\theta\,\cfrac{\pi_4^Q\,(\sigma_1+\pi_6^Q)}{\pi_7^Q}=\Gamma_{\bar\theta t}^{\theta}\\
\Gamma_{t\bar\theta}^{\bar\theta}&=&\Gamma_{t\bar\theta}^{\bar\theta\; CPI}(\pi_i^Q)-\bar\theta\theta\,\cfrac{\pi_6^Q\,(\pi_2^Q+\pi_3^Q)+2\,\pi_2^Q\,\sigma_1}{2\,\pi_7^Q}=\Gamma_{\bar\theta t}^{\bar\theta}\nonumber\\
\Gamma_{\theta\bar\theta}^{t}&=&\Gamma_{\theta\bar\theta}^{t\; CPI}(\pi_i^Q)-\bar\theta\theta\,\cfrac{(\pi_2^Q-\pi_3^Q)(\sigma_1+\pi_6^Q+2\,\pi_5^Q\pi_7^Q)}{2\,\pi_7^{Q^2}}\nonumber\\&=&-\Gamma_{\bar\theta \theta}^{t}\nonumber\\
\Gamma_{\theta\bar\theta}^{\theta}&=&\Gamma_{\theta\bar\theta}^{\theta\; CPI}(\pi_i^Q)-\theta\,\cfrac{\pi_6^Q}{\pi_7^Q}=-\Gamma_{\bar\theta\theta}^{\theta}\nonumber\\
\Gamma_{\theta\bar\theta}^{\bar\theta}&=&\Gamma_{\theta\bar\theta}^{\bar\theta\; CPI}(\pi_i^Q)-\bar\theta\,\cfrac{\pi_6^Q}{\pi_7^Q}=-\Gamma_{\bar\theta\theta}^{\bar\theta},\nonumber
\label{eq:B-2-1}
\end{eqnarray}
where 
\[\Gamma_{AB}^{C\; CPI}(\pi_i^Q)\equiv\Gamma_{AB}^{C\; CPI}(\pi_i\to\pi_i^Q,a\to \pi_7^Q).
\]
All the other Christofel symbols are equal to zero.
\par If we choose the $\pi_i$ dependent on time for the CPI Christofel symbols we get the following expressions: 
\begin{eqnarray}
\Gamma_{tt}^t&=&\bar\theta\theta\,\pi_5'\nonumber\\
\Gamma_{tt}^\theta&=&\theta\,\pi_3'+\bar\theta\,\pi_4'\nonumber\\
\Gamma_{tt}^{\bar\theta}&=&-\theta\,\pi_1'-\bar\theta\,\pi_2'\nonumber\\
\Gamma_{t\theta}^{t}&=&-\theta\,\frac{\pi_1\,(\pi_2-\pi_3)}{2\,a}+\bar\theta\,\frac{(\pi_2+\pi_3)\pi_2-2\,\pi_1\,\pi_4}{2\,a}\nonumber\\
\Gamma_{t\theta}^{\theta}&=&\frac{\pi_2+\pi_3}{2}+\bar\theta\theta\,\frac{\pi_5'}{2}\nonumber\\
\Gamma_{t\theta}^{\bar\theta}&=&-\pi_1\nonumber\\
\Gamma_{t\bar\theta}^{t}&=&\bar\theta\,\frac{(\pi_2-\pi_3)\pi_4}{2\,a}+\theta\,\frac{2\,\pi_1\,\pi_4-\pi_3\,(\pi_2+\pi_3)}{2\,a}\nonumber\\
\Gamma_{t\bar\theta}^{\theta}&=&\pi_4\nonumber\\
\Gamma_{t\bar\theta}^{\bar\theta}&=&-\frac{\pi_7}{2}+\bar\theta\theta\,\frac{\pi_5'}{2}\nonumber\\
\Gamma_{\theta t}^{t}&=&-\Gamma_{t\theta}^t\nonumber\\
\Gamma_{\theta t}^{\theta}&=&\Gamma_{t\theta}^{\theta}\nonumber\\
\Gamma_{\theta t}^{\bar\theta}&=&\Gamma_{t\theta}^{\bar\theta}\nonumber\\
\Gamma_{\theta\bar\theta}^t&=&\frac{\pi_2-\pi_3}{2\,a}-\bar\theta\theta\,\frac{4\,\pi_2\,\pi_5-\pi_5'}{2\,a}\nonumber\\
\Gamma_{\theta\bar\theta}^{\theta}&=&\bar\theta\,\frac{(\pi_2-\pi_3)\,\pi_4}{2\,a}+\theta\,\frac{(\pi_2-\pi_3)\,\pi_3+2\,a\,\pi_5}{2\,a}\nonumber\\
\Gamma_{\theta\bar\theta}^{\bar\theta}&=&-\bar\theta\,\frac{(\pi_2-\pi_3)\,\pi_1}{2\,a}+\bar\theta\,\frac{2\,a\,\pi_5-(\pi_2-\pi_3)\,\pi_2}{2\,a}\nonumber\\
\Gamma_{\bar\theta t}^{t}&=&-\Gamma_{t\bar\theta}^t\nonumber\\
\Gamma_{\bar\theta t}^{\theta}&=&\Gamma_{t\bar\theta}^{\theta}\nonumber\\
\Gamma_{\bar\theta t}^{\bar\theta}&=&-\frac{\pi_2+\pi_3}{2}+\bar\theta\theta\,\frac{\pi_2'+\pi_3'}{2}\nonumber\\
\Gamma_{\bar\theta \theta}^{t}&=&-\Gamma_{\theta\bar\theta}^t\nonumber\\
\Gamma_{\bar\theta \theta}^{\theta}&=&-\Gamma_{\theta\bar\theta}^\theta.\nonumber\\
\Gamma_{t\theta}^{\theta}&=&-\frac{\pi_2+\pi_3}{2}+\bar\theta\theta\,\frac{\pi_5'}{2}\nonumber\\
\end{eqnarray}
For the QPI case, when the coefficient $\pi_i$ depend on time the Christofel symbols turn out to be:
\begin{eqnarray}
\Gamma_{tt}^t&=&\Gamma_{tt}^{t\;CPI}-\bar\theta\theta\,\frac{(\pi_2-\pi_3)\,\sigma_1}{\pi_7}\nonumber\\
\Gamma_{tt}^\theta&=&\Gamma_{tt}^{\theta\;CPI}-\theta\,\sigma_1\nonumber\\
\Gamma_{tt}^{\bar\theta}&=&\Gamma_{tt}^{\bar\theta\;CPI}-\bar\theta\,\sigma_1\nonumber\\
\Gamma_{t\theta}^{t}&=&\Gamma_{t\theta}^{t\;CPI}+\bar\theta\,\frac{\sigma_1}{\pi_7}\nonumber\\
\Gamma_{t\theta}^{\theta}&=&\Gamma_{t\theta}^{\theta\;CPI}+\bar\theta\theta\,\frac{\sigma_1'+\pi_6\,(\pi_2+\pi_3)+2\,\pi_3\,\sigma_1-\pi_6'}{2\,\pi_7}\nonumber\\
\Gamma_{t\theta}^{\bar\theta}&=&\Gamma_{t\theta}^{\bar\theta\;CPI}-\bar\theta\theta\,\frac{\pi_1\,(\sigma_1+\pi_6)}{\pi_7}\nonumber\\
\Gamma_{t\bar\theta}^{t}&=&\Gamma_{t\bar\theta}^{t\;CPI}-\theta\,\frac{\sigma_1}{\pi_7}\nonumber\\
\Gamma_{t\bar\theta}^{\theta}&=&\Gamma_{t\bar\theta}^{\theta\;CPI}+\bar\theta\theta\,\frac{\pi_4\,(\sigma_1+\pi_6)}{\pi_7}\nonumber\\
\Gamma_{t\bar\theta}^{\bar\theta}&=&\Gamma_{t\bar\theta}^{\bar\theta\;CPI}+\bar\theta\theta\,\frac{\sigma_1'-\pi_6\,(\pi_2+\pi_3)-2\,\pi_2\,\sigma_1-\pi_6'}{\pi_7}\nonumber\\
\Gamma_{\theta t}^{t}&=&-\Gamma_{t\theta}^{t}\nonumber\\
\Gamma_{\theta t}^{\theta}&=&\Gamma_{t\theta}^{\theta}\nonumber\\
\Gamma_{\theta t}^{\bar\theta}&=&\Gamma_{t\theta}^{\bar\theta}\nonumber\\
\Gamma_{\theta \bar\theta}^{t}&=&\Gamma_{\theta\bar\theta}^{t\;CPI}-\bar\theta\theta\,\frac{\pi_6'-\sigma_1'+2\,(\pi_2-\pi_3)(\sigma_1-\pi_6)}{2\,\pi_7^2}\nonumber\\
\Gamma_{\theta \bar\theta}^{\theta}&=&\Gamma_{\theta\bar\theta}^{\theta\;CPI}+\theta\,\frac{\sigma_1-\pi_6}{\pi_7}\nonumber\\
\Gamma_{\theta \bar\theta}^{\bar\theta}&=&\Gamma_{\theta\bar\theta}^{\bar\theta\;CPI}+\bar\theta\,\frac{\sigma_1-\pi_6}{\pi_7}\nonumber\\
\Gamma_{\bar\theta t}^{t}&=&-\Gamma_{t \bar\theta}^{t}\nonumber\\
\Gamma_{\bar\theta t}^{\theta}&=&\Gamma_{t \bar\theta}^{\theta}\nonumber\\
\Gamma_{\bar\theta t}^{\bar\theta}&=&\Gamma_{\bar\theta t}^{\bar\theta}\nonumber\\
\Gamma_{\bar\theta \theta}^{t}&=&-\Gamma_{\theta \bar\theta}^{t}\nonumber\\
\Gamma_{\bar\theta \theta}^{\theta}&=&-\Gamma_{\theta \bar\theta}^{\theta}\nonumber\\
\Gamma_{\bar\theta \theta}^{\bar\theta}&=&-\Gamma_{\theta \bar\theta}^{\bar\theta}\nonumber.
\end{eqnarray}
The other symbols are zero.
%%%%%%%%%%%%%%%%%%%%%%%%%%%%%%%%%%%%%%%%%%%%%%%%
%%%%%%%%%%%%%%%%%%%%%%%%%%%%%%%%%%%%%%%%%%%%%%%%
%%%%%%%%%%%%%%%%%%%%%%%%%%%%%%%%%%%%%%%%%%%%%%%%
\newpage
\section{\label{Appendix:F}}
\subsection{Time Independent CPI Ricci tensor and scalar}
In the case of $\pi_i$ independent of $t$ the CPI  Ricci curvature tensor turns out to be: 
\begin{eqnarray}
\label{eq:C-1-1}
R_{\theta\theta}&=&R_{\bar\theta\bar\theta}=0\\
R_{tt}&=&\cfrac{(\pi_2+\pi_3)^2-4\,\pi_1\pi_4}{2}\nonumber\\
R_{t\theta}&=&-\cfrac{(\theta\,\pi_1+\bar\theta\,\pi_2)((\pi_2+\pi_3)^2-4\,\pi_1\pi_4)}{2\,a}=-R_{\theta t}\nonumber\\
R_{t\bar\theta}&=&-\cfrac{(\theta\,\pi_3+\bar\theta\,\pi_4)((\pi_2+\pi_3)^2-4\,\pi_1\pi_4)}{2\,a}=-R_{\bar\theta t}\nonumber\\
R_{\theta\bar\theta}&=&\cfrac{\bar\theta\theta}{2}(\pi_1\pi_4-\pi_2\pi_3)(\pi_2^2-6\,\pi_2\pi_3+\pi_3^2+4\,\pi_1\,\pi_4)\nonumber\\
&-&\cfrac{a}{2}\,(\pi_2^2-10\,\pi_2\pi_3+\pi_3^2+8\,\pi_1\,\pi_4)=-R_{\bar\theta\theta}\nonumber
\end{eqnarray}
From the components we can build the Ricci scalar
\begin{equation}
\label{eq:C-2-1}
R^{CPI}=-\cfrac{1}{2}\,(\pi_2^2-22\,\pi_2\pi_3+\pi_3^2+20\,\pi_1\,\pi_4)+8\,\cfrac{\bar\theta\theta}{a}\,(\pi_2\pi_3-\pi_1\pi_4)^2
\end{equation}
\subsection{Time Independent QPI Ricci tensor and scalar}
For the QPI, the components of the Ricci tensor can be written as: 
\begin{eqnarray}
\label{eq:C-3-1}
R_{\theta\theta}&=&R_{\bar\theta\bar\theta}=0\\
R_{tt}&=&R_{tt}^{CPI}(\pi_i^Q)+2\sigma_1+\cfrac{\bar\theta\theta}{\pi_7^Q}\left[\pi_6^Q((\pi_2^Q+\pi_3^Q)^2-4\,\pi_1^Q\pi_4^Q)+\right.\nonumber\\
&+&\left.2\,\sigma_1\,(\pi_2^{Q^2}+4\,\pi_2^Q\pi_3^Q+\pi_3^{Q^2}- 6\,\pi_1^Q\pi_4^Q-\pi_6^Q)+6\,\sigma_1^2\right]\nonumber\\
R_{t\theta}&=&R_{t\theta}^{CPI}(\pi_i^Q)-\cfrac{\pi_6^Q+3\,\sigma_1}{\pi_7^Q}\left[\theta\pi_1^Q+\bar\theta(\pi_2^Q+\pi_3^Q)\right]=-R_{\theta t}\nonumber\\
R_{t\bar\theta}&=&R_{t\bar\theta}^{CPI}(\pi_i^Q)-\cfrac{\pi_6^Q+3\,\sigma_1}{\pi_7^Q}\left[\bar\theta\pi_4^Q+\theta(\pi_2^Q+\pi_3^Q)\right]=-R_{\bar\theta t}\nonumber\\
R_{\theta\bar\theta}&=&R_{\theta\bar\theta}^{CPI}(\pi_i^Q)-\cfrac{\sigma_1-3\,\pi_6^Q}{\pi_7^Q}+\cfrac{\bar\theta\theta}{2\,\pi_7^{Q^2}}\left[2\,\pi_6^{Q^2}+8\,\pi_1^Q\pi_4^Q\pi_6^Q\right.\nonumber\\
&-&\left. 8\,\pi_2^Q\pi_3^Q\pi_6^Q-\sigma_1((\pi_2^Q-\pi_3^Q)^2+4\,\pi_6^Q+2\,\sigma_1)\right]=-R_{\bar\theta\theta},\nonumber
\end{eqnarray}
where 
\[R_{AB}^{CPI}(\pi_i^Q)\equiv R_{AB}^{CPI}(\pi_i\to\pi_i^Q,a\to \pi_7^Q).
\]
The Ricci scalar turns out to be 
\begin{eqnarray}
\label{eq:C-4-1}
R^{QPI}&=&R^{CPI}(\pi_i^Q)+2\,\sigma_1-3\,\pi_6^Q+\\
&+&\cfrac{\bar\theta\theta}{\pi_7^Q}\left[-\pi_6^Q(\pi_2^{Q^2}+6\,\pi_2^Q\pi_3^Q+\pi_3^{Q^2}-8\,\pi_1^Q\pi_4^Q)+4\,\pi_6^{Q^2}\right.
\nonumber\\
&-&\left. 4\,\sigma_1(\pi_2^{Q^2}+3\,\pi_2^Q\pi_3^Q+\pi_3^{Q^2}-5\,\pi_1^Q\pi_4^Q\-\pi_6^Q+\sigma_1)\right]\nonumber
\end{eqnarray}
where 
\[R^{CPI}(\pi_i^Q)=R^{CPI}(\pi_i\to\pi_i^Q,a\to \pi_7^Q).
\]
%%%%%%%%%%%%%%%%%%%%%%%%%%%%%%%%%%%%%%%%%%%%%%%%
%%%%%%%%%%%%%%%%%%%%%%%%%%%%%%%%%%%%%%%%%%%%%%%%
%%%%%%%%%%%%%%%%%%%%%%%%%%%%%%%%%%%%%%%%%%%%%%%%
\newpage
\section{\label{Appendix:G}}
\subsection{Time Dependent CPI Ricci tensor and scalar}
In the case of $\pi_i$ dependent on $t$ we get for the $CPI$ the expressions below where $\pi_i'$ and $\pi_i''$ indicate the first and second derivative of $\pi_i$ with respect to $t$.
\begin{align}
\label{eq:D-4-1}
R_{\theta\theta}&=R_{\bar\theta\bar\theta}=0\\
R_{tt}&=R_{tt}^{CPI}(\pi_i)+(\pi_3'-\pi_2')+\bar\theta\theta\left[\pi_3^2\pi_2'-\pi_2^2\pi_3'+\right.\nonumber\\
&+\left.(\pi_3'-\pi_2')(a\,\pi_5-2\,\pi_2\pi_3+3\,\pi_1\pi_4)+2\,a\pi_5'+\right.\nonumber\\
&+\left.(\pi_2-\pi_3)(\pi_4\pi_1'+\pi_4'\pi_1)\right]\nonumber\\
R_{t\theta}&=R_{t\theta}^{CPI}(\pi_i)+\cfrac{\theta\pi_1+\bar\theta\pi_2}{2\,a}(\pi_3'-\pi_2')+\cfrac{\bar\theta \pi_5'}{2}\nonumber\\
R_{t\bar\theta}&=R_{t\bar\theta}^{CPI}(\pi_i)+\cfrac{\theta\pi_3+\bar\theta\pi_4}{2\,a}(\pi_2'-\pi_3')-\cfrac{\theta \pi_5'}{2}\nonumber\\
R_{\theta t}&=R_{\theta t}^{CPI}(\pi_i)-\cfrac{\theta\pi_3+\bar\theta\pi_4}{2\,a}(\pi_2'-\pi_3')-\cfrac{3\,\theta \pi_5'}{2}\nonumber\\
R_{\theta \bar\theta}&=R_{\theta \bar\theta}^{CPI}(\pi_i)+\cfrac{a\,(\pi_2'-\pi_3')}{2}+\cfrac{\bar\theta\theta}{2}\left[5\,a\,\pi_5'(\pi_3-\pi_2)+\right.\nonumber\\
&+\left. 4\,a\,\pi_5(\pi_3'-\pi_2')+a\pi_5''\right]\nonumber\\
R_{\bar\theta t}&=R_{\bar\theta t}^{CPI}(\pi_i)-\cfrac{\theta\pi_3+\bar\theta\pi_4}{2\,a}(\pi_2'-\pi_3')-\cfrac{3\,\theta \pi_5'}{2}\nonumber\\
R_{\bar\theta\theta}&=R_{\bar\theta\theta }^{CPI}(\pi_i)-\cfrac{a\,(\pi_2'-\pi_3')}{2}-\cfrac{\bar\theta\theta}{2}\left[5\,a\,\pi_5'(\pi_3-\pi_2)+\right.\nonumber\\
&+\left. 4\,a\,\pi_5(\pi_3'-\pi_2')+a\pi_5''\right]\nonumber.
\end{align}
$R_{AB}^{CPI}(\pi_i)$ are the expressions previously presented in  
\ref{Appendix:C} for the time independent CPI Ricci tensor. The other tensor components are all equal to zero.
The CPI Ricci scalar, when the parameters are explicit functions of time, turns out to be:
\begin{eqnarray}
\label{eq:D-4-2}
R^{CPI}&=&R^{CPI}(\pi_i)+2(\pi_2'-\pi_3')+\\
&+&\bar\theta\theta\left[4\,\pi_5\,(\pi_3'-\pi_2')+7\,(\pi_3-\pi_2)\,\pi_5'+2\,\pi_5''\right],\nonumber
\end{eqnarray}
where, once again, $R^{CPI}(\pi_i)$ is the CPI Ricci scalar given in \ref{Appendix:C}.
\subsection{Time Dependent QPI Ricci Tensor and Curvature}
In the case of $\pi_i$ dependent on $t$ we get for the $QPI$ Ricci tensor:
\begin{eqnarray}
\label{eq:D-4-3}
R_{tt}&=&R_{tt}^{QPI}(\pi_i)+\pi _2'(t)-\pi _3'(t)+\\
&-&\cfrac{\bar\theta\theta}{\pi_7}\left[-\pi _2(t) \left(2 \pi _4(t) \pi _1'(t)+4 \pi _3(t) \left(\pi _2'(t)-\pi _3'(t)\right)+2 \pi _1(t)\pi _4'(t)+\right.\right.\nonumber\\
&+&\left.\left. \pi _7 \pi _5'(t)+\pi _3''(t)\right)+\pi _3(t) \left(2 \pi _4(t) \pi _1'(t)+2 \pi _1(t) \pi _4'(t)+\pi _7 \pi _5'(t)+\right.\right.\nonumber\\
&-&\left.\left.\left. \pi_2''(t)\right)+2 \pi _2(t){}^2 \pi _3'(t)-2 \pi _3(t){}^2 \pi _2'(t)+6 \pi _1(t) \pi _4(t) \pi _2'(t)+\right.\right.\nonumber\\
&+&\left.\left. 2 \pi _7 \pi _5(t) \pi_2'(t)+\pi _6(t) \pi _2'(t)-6 \pi _1(t) \pi _4(t) \pi _3'(t)-2 \pi _7 \pi _5(t) \pi _3'(t)+\right.\right.\nonumber\\
&-&\left.\left. \pi _6(t) \pi _3'(t)-2 \pi _2'(t) \pi_3'(t)+2 \pi _1'(t) \pi _4'(t)+\pi _4(t) \pi _1''(t)+\pi _1(t) \pi _4''(t)+\right.\right.\nonumber\\
&+&\left. \pi _6''(t)\right]\nonumber\\
R_{t\theta}&=&R_{t\theta}^{QPI}(\pi_i)+\cfrac{\theta }{2 \pi _7}\,\left(\pi _1(t) \pi _2'(t)-\pi _1(t) \pi _3'(t)\right)+\frac{\bar\theta}{2 \pi _7}\, \left[-\pi _4(t) \pi _1'(t)+\right.\nonumber\\
&+&\left.\left(\pi_2(t)+\pi _3(t)\right) \pi _2'(t)-\pi _1(t) \pi _4'(t)+3 \pi _6'(t)\right]\nonumber\\
R_{t\bar\theta}&=&R_{t\bar\theta}^{QPI}(\pi_i)+\frac{\theta}{2 \pi _7}\left[\pi _4(t) \pi _1'(t)-\left(\pi _2(t)+\pi _3(t)\right) \pi _3'(t)+\pi _1(t) \pi _4'(t)+\right.\nonumber\\
&+&\left. -3 \pi _6'(t)\right]+\frac{\bar\theta}{2\,\pi_7}\,\left(\pi _4(t)   \left(\pi _2'(t)-\pi _3'(t)\right)\right)\nonumber\\
R_{\theta t}&=&R_{\theta t}^{QPI}(\pi_i)+\frac{\theta}{2 \pi _7}\,\pi _1(t) \left(\pi _3'(t)-\pi _2'(t)\right)+\frac{\bar\theta}{2 \pi _7}  \left[-3 \pi _4(t) \pi _1'(t)+\right.\nonumber\\
&+&\left.-\left(\pi _2(t)-3 \pi_3(t)\right) \pi _2'(t)+4 \pi _2(t) \pi _3'(t)-3 \left(\pi _1(t) \pi _4'(t)+\pi _6'(t)\right)\right]\nonumber\\
R_{\theta \bar\theta}&=&R_{\theta \bar\theta}^{QPI}(\pi_i)+\frac{\pi _2'(t)-\pi _3'(t)}{2 \pi _7}-\frac{\bar\theta\theta}{2\,\pi_7^2} \left[-\pi _2(t) \left(2 \pi _4(t) \pi _1'(t)+\right.\right. \nonumber \\
&+&\left.\left.4 \pi _3(t) \left(\pi _3'(t)-\pi _2'(t)\right)+2 \pi _1(t)\pi _4'(t)-3 \pi _7 \pi _5'(t)+2 \pi _6'(t)+\pi _3''(t)\right)+\right.\nonumber\\
&+&\left. \pi _3(t) \left(2 \pi _4(t) \pi _1'(t)+2 \pi _1(t) \pi _4'(t)-3 \pi_7 \pi _5'(t)+2 \pi _6'(t)-\pi _2''(t)\right)+\right.\nonumber\\
&+&\left. 2 \pi _2(t){}^2 \pi _3'(t)-2 \pi _3(t){}^2 \pi _2'(t)-2 \pi _1(t) \pi _4(t) \pi
   _2'(t)+2 \pi _7 \pi _5(t) \pi _2'(t)+\right.\nonumber\\
&-&\left. 2 \pi _6(t) \pi _2'(t)+2 \pi _1(t) \pi _4(t) \pi _3'(t)-2 \pi _7 \pi _5(t) \pi _3'(t)+2 \pi_6(t) \pi _3'(t)+\right.\nonumber\\
&-&\left. 2 \pi _2'(t) \pi _3'(t)+2 \pi _1'(t) \pi _4'(t)+\pi _4(t) \pi _1''(t)+\pi _1(t) \pi _4''(t)+\pi _6''(t)\right]\nonumber
\end{eqnarray}
\begin{eqnarray}
R_{\bar\theta t}&=&R_{\bar\theta t}^{QPI}(\pi_i)+\frac{\theta}{2\,\pi_7}\left[3 \pi _4(t) \pi _1'(t)+\pi _3(t) \left(\pi _3'(t)-4 \pi _2'(t)\right)+3\left(-\pi _2(t) \pi _3'(t)\right.\right.\nonumber\\
&+&\left.\left. +\pi _1(t) \pi_4'(t)+\pi _6'(t)\right)\right]+\frac{\bar\theta}{2 \pi _7} \pi _4(t)  \left[\pi _3'(t)-\pi _2'(t)\right]\nonumber\\
R_{\bar\theta \theta}&=&R_{\bar\theta \theta}^{QPI}(\pi_i)+\frac{\pi _3'(t)-\pi _2'(t)}{2 \pi _7}-\frac{\bar\theta\theta}{\pi_7^2}\left[\pi _2(t) \left(2 \pi _4(t) \pi _1'(t)+4 \pi _3(t)\left(\pi _3'(t)-\pi _2'(t)\right)\right.\right.\nonumber\\
&+&\left.\left. 2 \pi _1(t) \pi_4'(t)-3 \pi _7 \pi _5'(t)+2 \pi _6'(t)+\pi _3''(t)\right)-\pi _3(t) \left(2 \pi _4(t) \pi _1'(t)+\right.\right.\nonumber\\
&+&\left.\left. 2 \pi _1(t) \pi _4'(t)-3 \pi _7\pi _5'(t)+2 \pi _6'(t)-\pi _2''(t)\right)-2 \pi _2(t){}^2 \pi _3'(t)2 \pi _3(t){}^2 \pi _2'(t)\right.+\nonumber\\
&+&\left.\left. 2 \pi _1(t) \pi _4(t) \pi _2'(t)-2 \pi _7 \pi _5(t) \pi _2'(t)+2 \pi _6(t) \pi _2'(t)-2 \pi _1(t) \pi _4(t) \pi _3'(t)+\right.\right.\nonumber\\
&+&\left.\left. 2 \pi _7 \pi _5(t) \pi _3'(t)-2 \pi _6(t) \pi_3'(t)+2 \pi _2'(t) \pi _3'(t)-2 \pi _1'(t) \pi _4'(t)-\pi _4(t) \pi _1''(t)+\right.\right.\nonumber\\
&-&\left.\left.\pi _1(t) \pi _4''(t)-\pi _6''(t)\right]\right.\nonumber
\end{eqnarray}
where $R_{AB}^{CPI}(\pi_i)$ are the expression previously presented in  
\ref{Appendix:C} for the QPI Ricci tensor. The other tensor components are all equal to zero.
The QPI Ricci scalar, when the parameters are explicit functions of time, turns out:
\begin{eqnarray}
\label{eq:D-4-4}
R^{QPI}&=&R^{QPI}(\pi_i)+2(\pi_2'-\pi_3')+\\
&+&\cfrac{\bar\theta\theta}{\pi_7}\left[-\pi _2(t) \left(5 \pi _4(t) \pi _1'(t)+3 \pi _3(t) \left(\pi _3'(t)-\pi _2'(t)\right)+5 \pi _1(t) \pi _4'(t)\right.\right.\nonumber\\
&-&\left.\left. 2 \pi _7 \pi _5'(t)+5 \pi _6'(t)+2 \pi _3''(t)\right)+\pi _3(t) \left(5 \pi _4(t) \pi _1'(t)+5 \pi _1(t) \pi _4'(t)\right.\right.\nonumber\\
&-&\left.\left.2 \pi_7 \pi _5'(t)+5 \pi _6'(t)-2 \pi _2''(t)\right)+2 \left(\pi _1(t) \left(\pi _4(t) \left(\pi _2'(t)-\pi _3'(t)\right)+\pi
   _4''(t)\right)\right.\right.\nonumber\\
&+&\left.\left. \left(3 \pi _7 \pi _5(t)-\pi _6(t)\right) \pi _2'(t)+\left(-2 \pi _2'(t)-3 \pi _7 \pi _5(t)+\pi _6(t)\right)\pi _3'(t)\right.\right.\nonumber\\   
&+&\left.\left.  2 \pi _1'(t) \pi _4'(t)+\pi _4(t) \pi _1''(t)+\pi _6''(t)\right)+5 \pi _2(t){}^2 \pi _3'(t)-5 \pi _3(t){}^2 \pi_2'(t)\right]\nonumber
\end{eqnarray}
where, once again, $R^{QPI}(\pi_i)$ is the QPI Ricci scalar given in \ref{Appendix:C}. 
%%%%%%%%%%%%%%%%%%%%%%%%%%%%%%%%%%%%%%%%%%%%%%%%
%%%%%%%%%%%%%%%%%%%%%%%%%%%%%%%%%%%%%%%%%%%%%%%%
%%%%%%%%%%%%%%%%%%%%%%%%%%%%%%%%%%%%%%%%%%%%%%%%
\newpage
\section*{References}
\bibliography{mybibfile}

\end{document}